\documentclass[reprint,superscriptaddress,amsmath,amsfonts,amssymb,aps,prb,showkeys]{revtex4-2}
\usepackage{tabularx}
\usepackage{graphicx}
\usepackage{dcolumn}
\usepackage{bm, color}
\usepackage{amsmath}
\usepackage{amsfonts}
\usepackage{amssymb}
\usepackage{amsmath}
\usepackage{here}
\usepackage{multirow}
\usepackage{ulem}
\usepackage{cancel}
\providecommand{\U}[1]{\protect\rule{.1in}{.1in}}

\newcommand{\vv}[1]{\boldsymbol #1}%

\usepackage{hyperref}
\hypersetup{
	colorlinks = true,
}
\usepackage{physics}

\begin{document}

\title{Higher-order topological phases for time-reversal-symmetry breaking superconductivity in UTe$_2$}

\author{Yuki Yamazaki}
\affiliation{Condensed Matter Theory Laboratory, RIKEN CPR, Wako, Saitama 351-0198, Japan}
\author{Shingo Kobayashi}
\affiliation{RIKEN Center for Emergent Matter Science, Wako, Saitama 351-0198, Japan}
\date{\today}

\date{\today}

\begin{abstract}
The recent discovery of heavy-fermion superconductor UTe$_2$ has broadened the possibility of realizing exotic time-reversal-symmetry-breaking superconductivity. However, a comprehensive understanding of the topological phases in the superconducting states of UTe$_2$ is still lacking.  
Here, we present an exhaustive classification of topological phases for all time-reversal symmetry breaking pairing symmetries of UTe$_2$. Using the K theoretical classification approach, we uncover that 25 out of 36 possible pairing states are classified as higher-order topological phases, with some demonstrating hybrid-order topology through an intricate interplay of hinge and corner states. Furthermore, under the weak-coupling condition of the pair potentials, the possible pairing symmetries are constrained to $B_{ju} + i B_{ku}$, $A_{u} + i B_{j u}$, and $B_{j g} + iA_u$ ($j,k = 1,2,3$; $j \neq k$), where these symbols denote the irreducible representations of the point group $D_{2h}$. For these pairing states, the topological invariants are related to the Fermi surface topology via the Fermi-surface formula, enabling us to systematically diagnose higher-order topological phases. Using a tight-binding model, we demonstrate the higher-order topological phases of the mixed-parity $A_u + iB_{1g}$ superconductors, where the second-order and hybrid-order topological phases emerge as the number of Fermi surfaces enclosing the time-reversal invariant momentum evolves from two to four. The findings suggest that UTe$_2$ serves as a compelling platform for exploring higher-order topological superconductors with diverse topological surface states.
\end{abstract} 

\maketitle
      
\makeatletter
\def\ext@table{}
\makeatother
\makeatletter
\def\ext@figure{}
\makeatother

\section{Introduction}\label{sec: Introduction}
The pursuit of topological superconductivity remains a central challenge in condensed matter physics~\cite{Qi-Zhang2011,Tanaka12,SatoFujimoto16,Sato17, Haim2019a,Mizushima022001,Chiu2016}. Odd-parity superconductors, both with and without time-reversal (TR) symmetry, are of particular interest due to their intrinsic sign change or phase winding in the order parameter, which gives rise to topologically nontrivial states~\cite{Sato214526, Fu097001, Sato220504, Sato224511}. Examples include the superfluid $^3$He-B and A phases~\cite{Volovik03}, where the former hosts two-dimensional (2D) Majorana quasiparticles~\cite{Qi2009}, while the latter exhibits one-dimensional (1D) chiral Majorana edge states~\cite{Read10267,Kallin054502}. Identifying materials that support these properties and understanding their response to electromagnetic fields are promising research directions, which are of particular relevance for quantum computing applications that exploit the non-Abelian statistic of Majorana quasiparticles~\cite{Nayak2008} and for the determination of pairing symmetries through their electromagnetic response~\cite{xiong17,kobayashi2019, yamazaki043703,yamazaki094508,kobayashi224504,yamazaki073701,Kobayashi2024,Yamazaki2024}.

The heavy fermion superconductor UTe$_2$~\cite{Ran684,aoki2022} has emerged as a strong candidate for odd-parity spin-triplet superconductivity in a solid-state system. Several experimental observations support this possibility, including upper critical fields exceeding the Pauli limit~\cite{aoki2019unconventional,nakamine2019,knebel2019,ran2019extreme}, re-entrant superconductivity in high magnetic fields~\cite{knebel2019,ran2019extreme,Kinjo2023}, a reduction of the NMR Knight shift below the superconducting transition temperature~\cite{Ran684,nakamine2019,matsumura063701,Suetsugu2024}, and a zero-bias conductance peak~\cite{yoon2024probing, Qiangqiang2501}. 
Additional studies have suggested time-reversal-symmetry-breaking (TRSB) superconductivity, as indicated by the Kerr-effect measurements~\cite{Hayes2021}, the presence of point nodes in specific heat, thermal current, and penetration depth measurements~\cite{metz2019,kittaka2020,Ishihara2966}, chiral in-gap states detected via scanning tunneling microscopy~\cite{Jiao523}, and the surface microwave impedance measurements~\cite{Bae2644, Arthur2502}. However, recent high-quality sample measurements have
found no evidence of spontaneous TRSB at ambient pressure~\cite{Cairns415602,Thomas224501,Rosa33,Girod121101,Ajeesh041019,theuss2024single,Suetsugu2024}.       

 In addition, UTe$_2$ exhibits multiple superconducting phases under pressure~\cite{Braithwaite2019,Aoki2020Multiple,Aoki2021Field,Kinjo2736,Vasina096501}, making it a compelling system for studying unconventional superconductivity. As pressure increases, the superconducting phase present at ambient pressure is gradually suppressed, while another superconducting phase emerges, indicating the coexistence of two distinct phases. At the critical pressure ($\simeq 1.5$ GPa), both superconducting phases disappear abruptly, coinciding with the onset of an antiferromagnetic phase~\cite{lin2020tuning,Thomas2020,Knebel2020,Ran2020Enhancement,Li20201Magnetic,Valiska2021,knafo2023}. This correlation suggests that antiferromagnetic fluctuations may play a role in stabilizing or suppressing superconductivity. The intricate phase diagram of UTe$_2$ has motivated extensive theoretical studies on multi-component superconductors~\cite{Ishizuka094504,Shishidou104504,Kanasugi39,Kitamura214513,Tei064516,Hakuno104509,Henrik054521}, proposing exotic TRSB superconductivity such as non-unitary-odd-parity and mixed-parity pairing states. However, fundamental questions remain regarding the pairing mechanism, the symmetry of Cooper pairs, and the possible topological superconductivity.

 From the topological perspective, UTe$_2$ is a promising candidate for topological superconductivity, supported by strong evidence of the odd-parity spin-triplet pairing. Early studies proposed two possible topological superconducting states: time-reversal-invariant topological superconductivity in class DIII~\cite{Ishizuka217001} and TRSB chiral superconductivity in class D~\cite{Jiao523,Shishidou104504,moriya2022}. These phases serve as solid-state analogs of the superfluid $^3$He-B and A phases, respectively.
 
The presence of three-dimensional (3D) Fermi surface is essential for strong topological superconductivity in class DIII. Experimental evidence, including  angle-resolved photoemission spectroscopy~\cite{fujimori2019,Miao076401} and quantum oscillation measurements~\cite{Broyles036501} supported a 3D Fermi surface.  However, recent investigations using the de Haas-van Alphen effect~\cite{Aoki083704}, magnetoconductance~\cite{Weinberger2024}, and magnetoresistance~\cite{Aoki123702} under a high magnetic field have revealed quasi-2D Fermi surfaces. These findings suggest that superconductivity in UTe$_2$ may not be topologically nontrivial in the sense of 3D class DIII.  
Based on these experimental results, another possibility of topological superconductivity stabilized under crystalline symmetry has been proposed~\cite{Tei144517,Henrik054521}. Recent theoretical studies of the electronic structure indicate that a 3D Fermi surface can be induced by electron correlation effects~\cite{Choi2024,Kang2025coexistence} or depends on the applied pressure~\cite{Shimizu2024}. To date, the discrepancy between the experimental findings remains unresolved, posing a challenge to the comprehensive understanding of possible topological phases and raising questions about how the electronic structure influences the topology of multi-component superconductors in class D.

We study the possible topological phases of the superconducting states of UTe$_2$, focusing on TRSB superconductors with multi-component order parameters. Our findings reveal that the presence of a 3D Fermi surface plays a crucial role in the realization of higher-order topological superconductivity in class D. The superconducting order parameters in UTe$_2$ transform according to the irreducible representations (IRs) of the crystalline point group, $mmm$ ($D_{2h}$). These order parameters are classified into $A_g$, $A_{jg}$, $A_u$, $B_{ju}$ ($j = 1,2,3$), along with their possible combinations. The symbols $A_g$, etc., denote the IRs of $mmm$, with subscript $g (u)$ indicating even (odd) parity. Assuming spontaneous TRSB, there are 36 possible pairing states, including non-unitary-odd-parity and mixed-parity pairings, such as $B_{2u} + iB_{3u}$ and $B_{1g} + i A_u$. These TRSB pairing states are not classified by the IRs of the point group but instead by the IRs of the magnetic point group, which comprises point groups, a TR operation, and their combinations. By integrating the symmetry classification of TRSB order parameters with the recently developed K-theoretical classification approach~\cite{Geier2018,Chapman075105,Luka011012,Luka202000090,Cornfeld2021,Shiozaki04A104}, we establish a complete topological classification of TRSB superconductivity in UTe$_2$. The classification predicts that 25 out of 36 pairing states can host higher-order topological phases (HOTPs), which manifest as 3D class D topological phases with Majorana hinge and/or corner states at crystal boundaries respecting magnetic point group symmetry. Using bulk topological invariants and boundary classification methods, we determine the possible configurations of topological surface states, where multiple Majorana hinge and corner states coexist in a complex manner. These surface states behave differently from those found in earlier studies on higher-order topology in TRSB superconductors~\cite{Hassan094508, Yuxuan165144, Akbar020509, Bitan220506, Bo180504, Bitan180503, Jahin053,Henrik054521}.    

Under the weak-coupling assumption that inter-band pairings are negligible, we derive the weak-coupling condition for HOTPs, which allows us to systematically identify nontrivial topological phases using information about the IRs of the Cooper pairings and the Fermi surface topology. In the weak-coupling limit, superconducting nodes emerge at intersections between the Fermi surface and high-symmetry lines or planes for certain pairing states~\cite{Sigrist239}, This constraint restricts the possible candidates to $B_{ju} + i B_{ku}$, $A_{u} + i B_{j u}$, and $B_{j g} + iA_u$ ($j,k = 1,2,3$; $j \neq k$). The corresponding topological invariants are calculated using a simple formula based on the number of Fermi surfaces enclosing time-reversal invariant momenta (TRIM) and the sign of pair potentials on the Fermi surface.

To illustrate HOTPs, we employ a tight-binding model with a mixed-parity $B_{1g}+iA_{u}$ pairing, where the topology depends on the number of Fermi surfaces (\#FS). 
By imposing full open boundary conditions compatible with the magnetic point group symmetry, we find that $\text{\#FS}=2$ yields the second-order topological phases characterized by Majorana hinge states on two orthogonal mirror planes and $\text{\#FS}=4$ corresponds to hybrid-order topological phases featuring both Majorana hinge and corner states. These results provide an insight into the intricate interplay of higher-order topology in TRSB superconductors.

\begin{figure*}[t] 
\centering
    \includegraphics[scale=0.049]{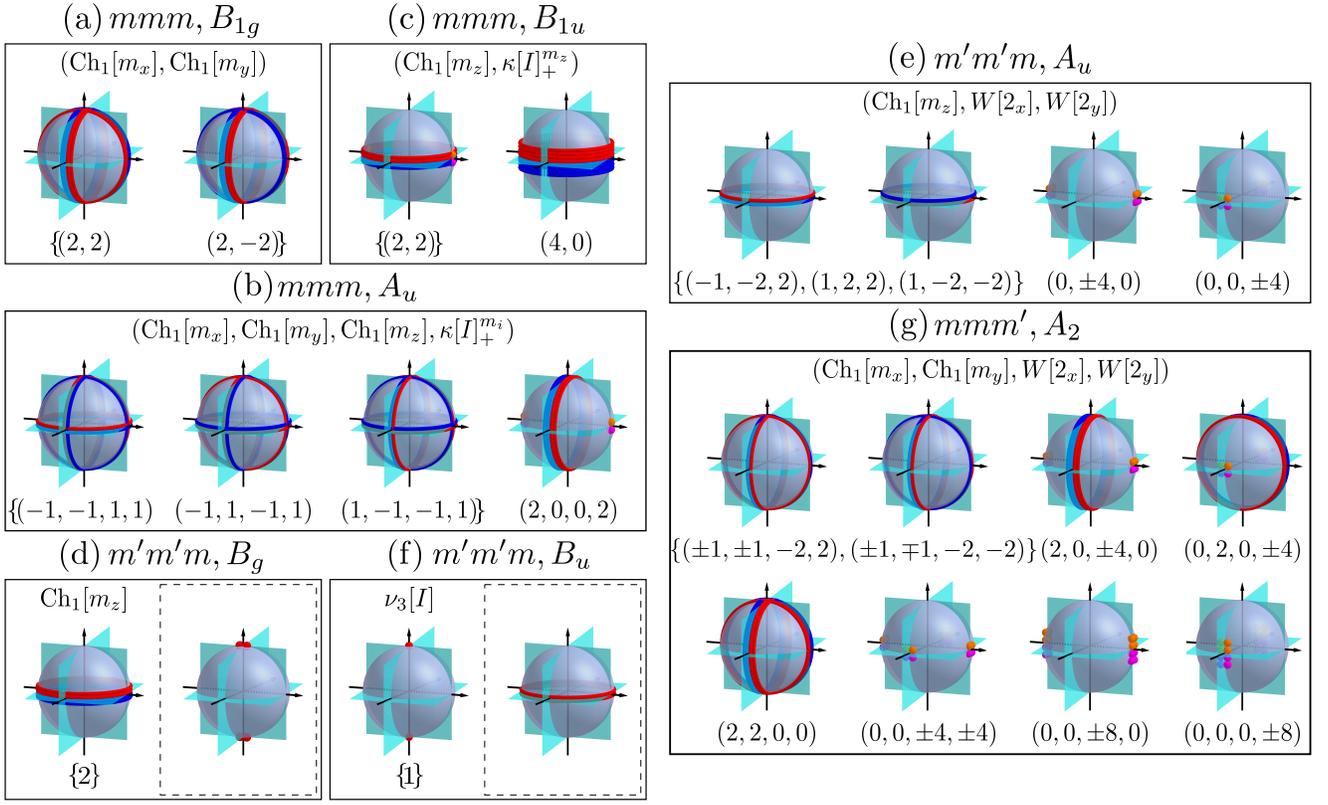}
    \caption{Figure panels illustrate possible topological surface states in 3D TRSB superconductors with pairing symmetry (a) $B_{1g}+i B_{1g}$, (b) $A_{u}+i A_{u}$, (c) $B_{1u}+iB_{1u}$, (d) $B_{2g}+i B_{3g}$, (e) $A_u+i B_{1u}$, (f)  $B_{2u}+i B_{3u}$, and (g) $B_{1g}+i A_{u}$. Each panel shows the topological invariants and configurations of topological surface states. The configurations are depicted by gray spheres, representing system boundaries, and colored lines and points, which indicate hinge and corner states, respectively. For instance, in case (a), two mirror Chern numbers, $\text{Ch}_1[m_x]$ and $\text{Ch}_1[m_y]$, define the HOTPs. The two surface state configurations shown correspond to topological invariants $\{(2,2),(2,-2)\}$, respectively. Here, the set notation $\{\}$ denotes a minimal set of topological invariants (see Appendix~\ref{app:generator}). Elements of the classifying groups are generated by the combinations of these minimal sets. In case (c), two topological invariants, $\text{Ch}_1[m_z]$ and $\kappa[I]_+^{m_z}$, correspond to second and third order topological phases, respectively. The minimal set \{$(2,2)$\} indicates coexistence of Majorana hinge and corner states. Doubling this set yields $(4,4)=(4,0)$, because $\kappa[I]_+^{m_z} = 4 =0 \mod 4$. This phase therefore supports only four Majorana hinge states. In cases (b), (e), and (g), diverse surface state patterns result from the coexistence of multiple topological invariants. In cases (d) and (f), an alternative configuration shown in the dotted box is possible through the addition of extrinsic topological surface states (see Appendix~\ref{app:boundary}).}
    \label{Fig:Boundary-states}
\end{figure*}

This paper is organized as follows. Section~\ref{sec:pair_symmetry} presents the symmetry classification of multi-component order parameters under magnetic point groups. Section~\ref{sec:topo_phase} provides a classification of all possible HOTPs in TRSB superconductors. The K theoretical classification and the topological invariants relevant to HOTPs are discussed in Sec.~\ref{sec:K-theory} and Sec.~\ref{sec:topo_inv}. The topological classification and the configuration of the topological surface states are summarized in Table~\ref{tab:pairings} and Figure~\ref{Fig:Boundary-states}. Section~\ref{sec:wca} examines the weak-coupling condition for HOTPs. Section~\ref{sec:model} applies the theory to a tight-binding model with the $B_{1g} + iA_u$ pairing and demonstrate the appearance of various topological surface states. The appendices include the bulk classification in Appendix~\ref{app:bulk}, the boundary classification in Appendix~\ref{app:boundary}, the definition of the topological invariants in Appendix~\ref{app:invariant}, and the relationship between the topological invariants in Appendix~\ref{app:generator}.

\section{Symmetry of multi-component order parameters}
\label{sec:pair_symmetry}
\begin{table}
	\caption{Irreducible representations (IRs) of $mmm$, where $E$ represents the identity operation, $I$ the spatial inversion, $2_i$ the twofold rotation around $i$ axis, and $m_{i}$ the mirror-reflection symmetry in terms of the plane normal to the $i$ axis.}
    \label{tab:ir-mmm}
		\begin{tabular}{ccccccccc}
            \hline \hline
            IR & $E$ & $2_z$ & $2_y$ & $2_x$ & $I$ & $m_{z}$ & $m_{y}$ & $m_{x}$ \\ \hline
            $A_g$ & $1$ & $1$ & $1$ & $1$ & $1$ & $1$ & $1$ & $1$  \\
            $B_{1g}$ & $1$ & $1$ & $-1$ & $-1$ & $1$ & $1$ & $-1$ & $-1$  \\
            $B_{2g}$ & $1$ & $-1$ & $1$ & $-1$ & $1$ & $-1$ & $1$ & $-1$  \\
            $B_{3g}$ & $1$ & $-1$ & $-1$ & $1$ & $1$ & $-1$ & $-1$ & $1$ \\
            $A_u$ & $1$ & $1$ & $1$ & $1$ & $-1$ & $-1$ & $-1$ & $-1$  \\
            $B_{1u}$ & $1$ & $1$ & $-1$ & $-1$ & $-1$ & $-1$ & $1$ & $1$  \\
            $B_{2u}$ & $1$ & $-1$ & $1$ & $-1$ & $-1$ & $1$ & $-1$ & $1$  \\
            $B_{3u}$ & $1$ & $-1$ & $-1$ & $1$ & $-1$ & $1$ & $1$ & $-1$  \\
            \hline \hline
            \end{tabular}
\end{table}
We present the symmetry classification of the possible TRSB pairing states of UTe$_2$. We assume that TR symmetry ($T:T^2=-1$) is preserved in the normal state and spontaneously broken in the superconducting states. 
UTe$_2$ has a body-centered orthorhombic lattice with space-group symmetry $Immm$ (SG\# 71). The relevant point group is $mmm$, which consists of the symmetry operations
\begin{align}
    mmm=\{E,2_x,2_y,2_z,I,m_{x},m_{y},m_{z}\},
\end{align}
where $E$ represents the identity operation, $I$ the spatial inversion, $2_i$ the twofold rotation around the $i$ axis, and $m_{i}$ the mirror-reflection symmetry in terms of the plane normal to the $i$ axis. We adopt the Hermann-Mauguin notation of magnetic point groups. Here, $\{\cdots\}$ represents a set of generators. Cooper pairs are formed by Bloch functions that respect the crystal symmetry $mmm$. Thus, possible pairings under the point group symmetry $mmm$  are classified according to eight IRs of $mmm$, as shown in Table~\ref{tab:ir-mmm}. Since all IRs are one-dimensional, TRSB states arise from combinations of different IRs, leading to 36 possible pairing states. These include TRSB pairing with non-unitary odd-parity and mixed-parity pairings, such as $A_{u} + i B_{1u}$ and $A_g+iB_{1u}$. A multi-component order parameter emerges when two superconducting states with different IRs are accidentally degenerate.   

Although these states preserve neither $mmm$ nor TR symmetry, they remain invariant under magnetic point groups. The magnetic point group $M$ is formally represented as
\begin{align}
    M=H + T (G-H),
\end{align}
where $G=mmm$, $H \subseteq mmm$, and $T(G-H)$ is a set of the magnetic symmetry operations.
For example, consider the $A_u + i B_{1u}$ state, where the pairing state of the $A_u$ ($B_{1u}$) state is even (odd) under $T$, and the transformation under  $mmm$ symmetry operations follows its IR as described in Table~\ref{tab:ir-mmm}. The $A_u + i B_{1u}$ state belongs to an IR of magnetic point groups,
\begin{align}
    m'm'm = \{e,2_z, I, m_{z}; T2_x,T2_y,Tm_{x}, Tm_{y}\}, \label{eq:m'm'm}
\end{align}
where $H=112/m$, and we use the notation $M=\{H;T(G-H)\}$.
Since the $A_u + i B_{1u}$ state is even (odd) under $2_z$ ($I$ and $m_{xy}$), it belongs to the $A_u$ IR of $H=112/m$. As such, 36 TRSB pairing states can be assigned to one of the IRs of magnetic point groups (see Table~\ref{tab:pairings}), where $m'mm$ and $m'm'm'$ are defined by
\begin{align}
    mmm' &= \{e,2_z, m_{x}, m_{y}; TI, T2_x,T2_y,Tm_{z}\}, \label{eq:m'mm} \\
    m'm'm' &= \{e,2_x, 2_y, 2_z ; TI, Tm_x,Tm_y,Tm_{z}\}. \label{eq:m'm'm'}
\end{align}
The other groups are obtained by permuting $x,y,z$, e.g., $mm'm$ is given by permuting $(x,y,z) \to (y,z,x)$ in Eq.~(\ref{eq:m'mm}). Note that the IRs of $H$ are independent of the gauge choice of the pairing states. For example, $B_{1u} + iA_u$ and $A_u + i B_{1u}$ both belong to the $A_u$ IR of $H=112/m$.

\begin{table*}
	\caption{Classification of HOTPs for TRSB superconducting states of UTe$_2$. The first, second, third, and fourth columns show the relationship between magnetic point groups $M$, unitary subgroups of $M$ ($H \subseteq M$), IRs of $H$, corresponding multi-component order parameters. The fifth column classifies types of pairings as (i) unitary pairings, (ii) non-unitary pairings, or (iii) mixed-parity pairings. The sixth, seventh, and eighth columns represent the K theoretical classification of intrinsic surface states, where $\mathcal{K}_a^{(1)}$, $\mathcal{K}_a^{(2)}$, and $\mathcal{K}_a^{(3)}$ classify the first, second, and third order topological phases, respectively. The ninth column presents the topological invariants relevant to HOTPs. The last column shows the connection between the topological invariants and surface state configurations, which are illustrated in Figure~\ref{Fig:Boundary-states}.} \label{tab:pairings}
		\begin{tabular}{ccccccccccc}
            \hline \hline
		  $M$ & $H$ & IR  & Pairings & Type & $\mathcal{K}_a^{(1)}$ & $\mathcal{K}_a^{(2)}$ & $\mathcal{K}_a^{(3)} $  & Topological invariants & Figure
			\\
			\hline    
            $mmm$ & $mmm$ & $A_g$ & $A_{g}+i A_{g}$ & i & $0$ & $0$ &$0$ &
            \\
            $mmm$ & $mmm$ & $B_{1g}$ & $B_{1g}+i B_{1g}$ & i &  $0$ & $\mathbb{Z}^2$& $0$ & $\text{Ch}_1[m_{x}], \text{Ch}_1[m_{y}] $ & (a)
            \\
            $mmm$ & $mmm$ & $B_{2g}$ & $B_{2g}+i B_{2g}$ & i &  $0$ & $\mathbb{Z}^2$& $0$ &$\text{Ch}_1[m_{z}], \text{Ch}_1[m_{x}] $  & (a)
            \\
            $mmm$ & $mmm$ & $B_{3g}$ & $B_{3g}+i B_{3g}$  & i &  $0$ & $\mathbb{Z}^2$& $0$ &$\text{Ch}_1[m_{y}], \text{Ch}_1[m_{z}] $  & (a)
            \\
            $mmm$ & $mmm$ & $A_u$ & $A_{u}+i A_{u}$ & i &  $0$ & $\mathbb{Z}^3$ & $\mathbb{Z}_2$  &$  \text{Ch}_1[m_{x}],\text{Ch}_1[m_{y}],\text{Ch}_1[m_{z}], \nu_2 [I]^{m_i}_{+}$  & (b)
            \\
            $mmm$ & $mmm$ & $B_{1u}$ & $B_{1u}+iB_{1u}$ & i & $0$ & $\mathbb{Z}$ & $\mathbb{Z}_2$  &$ \text{Ch}_1[m_{z}], \nu_2 [I]^{m_z}_{+}$  & (c)
            \\
            $mmm$ & $mmm$ & $B_{2u}$ & $B_{2u}+iB_{2u}$ & i & $0$ & $\mathbb{Z}$ & $\mathbb{Z}_2$  &$\text{Ch}_1[m_{y}], \nu_2 [I]^{m_y}_{+}$  & (c)
            \\
            $mmm$ & $mmm$ & $B_{3u}$ & $B_{3u}+iB_{3u}$ & i & $0$ & $ \mathbb{Z}$ & $\mathbb{Z}_2$  &$\text{Ch}_1[m_{x}], \nu_2 [I]^{m_x}_{+}$  & (c)
             \\
             $m'm'm$ & $112/m$ & $A_g$ & $A_{g}+i B_{1g}$ & ii & $0$ & $0$ &$0$ &
            \\
            $m'm'm$ & $112/m$ & $B_g$ & $B_{2g}+i B_{3g}$ & ii& $0$ &$\mathbb{Z}$ & $0$ & $\text{Ch}_1[m_{z}]$ & (d)
            \\
            $m'm'm$ & $112/m$ & $A_u$ & $A_u+i B_{1u}$ & ii& $0$ &$\mathbb{Z}$ & $\mathbb{Z}^2$  & $\text{Ch}_1[m_{z}], W[2_x],W[2_y]$  & (e)
            \\
            $m'm'm$ & $112/m$ & $B_u$ & $B_{2u}+i B_{3u}$ & ii & $0$  & $0$ & $\mathbb{Z}_2$  & $\nu_3[I]$  & (f)
            \\
            $m'mm'$ & $12/m1$ & $A_g$ & $A_{g}+i B_{2g}$ & ii& $0$ & $0$ &$0$ &
            \\
            $m'mm'$ & $12/m1$ & $B_g$ & $B_{1g}+i B_{3g}$ & ii& $0$ &$\mathbb{Z}$ & $0$  & $\text{Ch}_1[m_{y}]$  & (d)
            \\
            $m'mm'$ & $12/m1$ & $A_u$ & $A_u+i B_{2u}$ & ii& $0$ &$\mathbb{Z}$ & $\mathbb{Z}^2$  &$\text{Ch}_1[m_{y}],W[2_z],W[2_x]$  & (e)
            \\
            $m'mm'$ & $12/m1$ & $B_u$ & $B_{1u}+i B_{3u}$ & ii& $0$  & $0$ & $\mathbb{Z}_2$   & $\nu_3[I]$  & (f)
            \\
            $mm'm'$ & $2/m11$ & $A_g$ & $A_{g}+i B_{3g}$ & ii& $0$ & $0$ &$0$ &
            \\
            $mm'm'$ & $2/m11$ & $B_g$ & $B_{1g}+i B_{2g}$ & ii& $0$ &$\mathbb{Z}$ & $0$ & $\text{Ch}_1[m_{x}]$  & (d)
            \\
            $mm'm'$ & $2/m11$ & $A_u$ & $A_u+i B_{3u}$ & ii& $0$ &$\mathbb{Z}$ & $\mathbb{Z}^2$ &$ \text{Ch}_1[m_{x}],W[2_y],W[2_z]$  & (e)
            \\
            $mm'm'$ & $2/m11$ & $B_u$ & $B_{1u}+i B_{2u}$ & ii& $0$  & $0$ & $\mathbb{Z}_2$  & $\nu_3[I]$  & (f)
            \\
            $mmm'$ & $mm2$ & $A_1$ & $A_{g}+i B_{1u}$ & iii& $0$ & $0$ &$0$ &
            \\
            $mmm'$ & $mm2$ & $A_2$ & $B_{1g}+i A_{u}$ &iii& $0$ &$\mathbb{Z}^2$ & $\mathbb{Z}^2$ &$\text{Ch}_1[m_{x}],\text{Ch}_1[m_{y}],W[2_x],W[2_y]$  & (g)
            \\
            $mmm'$ & $mm2$ & $B_1$ & $B_{2g}+i B_{3u}$ & iii& $0$ &$\mathbb{Z}$ & $0$  & $\text{Ch}_1[m_{x}]$  & (d)
            \\ 
            $mmm'$ & $mm2$ & $B_2$ & $B_{3g}+i B_{2u}$ &iii &$0$ &$\mathbb{Z}$ & $0$  & $\text{Ch}_1[m_{y}]$  & (d)
            \\
            $mm'm$ & $m2m$ & $A_1$ & $A_{g}+i B_{2u}$ &iii& $0$ & $0$ &$0$ &
            \\
            $mm'm$ & $m2m$ & $A_2$ & $B_{2g}+i A_{u}$ &iii& $0$ &$\mathbb{Z}^2$ & $\mathbb{Z}^2$ &$\text{Ch}_1[m_{z}],\text{Ch}_1[m_{x}],W[2_z],W[2_x]$ & (g)
            \\
            $mm'm$ & $m2m$ & $B_1$ & $B_{1g}+i B_{3u}$ &iii& $0$ &$\mathbb{Z}$ & $0$  & $\text{Ch}_1[m_{x}]$  & (d)
            \\ 
            $mm'm$ & $m2m$ & $B_2$ & $B_{3g}+i B_{1u}$ &iii& $0$ &$\mathbb{Z}$ & $0$  & $\text{Ch}_1[m_{z}]$  & (d)
            \\
            $m'mm$ & $2mm$ & $A_1$ & $A_{g}+i B_{3u}$ & iii& $0$ & $0$ &$0$ &
            \\
            $m'mm$ & $2mm$ & $A_2$ & $B_{3g}+i A_{u}$ & iii& $0$ &$\mathbb{Z}^2$ & $\mathbb{Z}^2$ & $\text{Ch}_1[m_{y}],\text{Ch}_1[m_{z}],W[2_y],W[2_z]$  & (g)
            \\
            $m'mm$ & $2mm$ & $B_1$ & $B_{1g}+i B_{2u}$ & iii& $0$ &$\mathbb{Z}$ & $0$  & $\text{Ch}_1[m_{y}]$  & (d)
            \\ 
            $m'mm$ & $2mm$ & $B_2$ & $B_{2g}+i B_{1u}$ &iii& $0$ &$\mathbb{Z}$ & $0$  & $\text{Ch}_1[m_{z}]$ & (d)
            \\ 
            $m'm'm'$ & $222$ & $A$ & $A_{g}+iA_{u}$ & iii &$0$ & $0$ &$0$ &
            \\
            $m'm'm'$ & $222$ & $B_1$ & $B_{1g}+iB_{1u}$ &iii& $0$ & $0$ &$0$ &
            \\
            $m'm'm'$ & $222$ & $B_2$ & $B_{2g}+iB_{2u}$ &iii& $0$ & $0$ &$0$ &
            \\
            $m'm'm'$ & $222$ & $B_3$ & $B_{3g}+iB_{3u}$ &iii& $0$ & $0$ &$0$ &
		  \\
            \hline \hline
 \end{tabular}	
\end{table*}

The classification of order parameters under magnetic point groups serves as the basis for the topological classification. These order parameters fall into three distinct categories: (i) unitary pairings, (ii) non-unitary parings, and (iii) mixed-parity pairings. Here, the terms “unitary” and “non-unitary” are named from the unitarity of odd-parity pair potentials~\cite{Sigrist239}. In case (i), only TR symmetry is broken. Since $M=H=mmm$, no anti-unitary symmetry is present. In case (ii), spatial-inversion symmetry is preserved, ensuring a well-defined parity for the pairing states. The category includes non-unitary odd-parity chiral superconducting states~\cite{Shishidou104504,moriya2022,Choi2024}. In case (iii), both inversion and TR symmetries is broken, but their combination is preserved. Consequently, the Chern number is always zero, preventing the formation of a Weyl superconducting phase. This category includes anapole superconducting states~\cite{Kanasugi39,Kitamura214513}. 

\section{HOTPs for TRSB superconductors}
\label{sec:topo_phase}
In this section, we present a comprehensive classification of HOTPs for TRSB pairing states of UTe$_2$. TRSB superconductors belong to class D in the Altland-Zirnbauer symmetry classes~\cite{Altland1142}, which implies the absence of 3D topological invariants. Thus, 3D topological phases are solely characterized by lower-dimensional topological invariants, such as the Chern number and crystalline-symmetry-protected topological invariants. These topological phases are manifest as nodal superconducting phases with Majorana flat bands and higher-order topological phases featuring Majorana hinge and corner states. In contrast to the previous research on the Weyl superconducting phases of UTe$_2$~\cite{Jiao523,Shishidou104504,moriya2022}, the purpose of this study is to explore possible HOTPs that arise under the magnetic point group symmetry in the presence of multi-component order parameters and predict topological properties of UTe$_2$ that have been overlooked.

\subsection{Classification of possible HOTPs}
\label{sec:K-theory}
In 3D superconductors protected by the crystalline symmetry, HOTPs with lower-dimensional surface Majorana states emerge when the crystal faces and lattice termination are compatible with the crystalline symmetry. The order $n$ of a topological phases corresponds to the codimension of its boundary states, where $n=1$ represents a surface state, $n=2$ a hinge state, and $n=3$ a corner state.   
For HOTPs, surface states are categorized as extrinsic and intrinsic. Extrinsic topological surface states do not originate from bulk topology and can be removed without closing the bulk energy gap. In contrast, intrinsic topological surface states result from nontrivial bulk topology, which remains robust as long as the bulk energy gap is maintained and crystalline symmetry is preserved.

We classify possible intrinsic topological surface states using bulk and boundary classification approaches~\cite{Luka011012, Luka202000090}. From the bulk perspective, HOTPs are classified in the form of a subgroup series of the classifying groups,
\begin{align}
    K^{(3)} \subseteq K^{(2)} \subseteq K^{(1)} \subseteq K, \label{eq:k-series}
\end{align}
where $K$ is the classifying group that describes the 3D superconductors preserving the magnetic point group symmetry~\cite{shiozaki14,Shiozaki04A104}. In the absence of the crystalline symmetries, $K$ corresponds to the classifying group in the ten-fold way classification. For instance, 3D TRSB superconductors in class D satisfy $K=0$. The subgroups, $K^{(n)} \subseteq K$, are the classifying groups that exclude topological phases of order $n$ or lower for any crystal shape that preserve the underlying crystal symmetry. The subgroups are determined by  encoding information about the magnetic point group and the IRs of pair potentials and applying the Cornfeld-Chapman isomorphism~\cite{Chapman075105} to an effective Dirac Hamiltonian. See Appendix~\ref{app:bulk} for further details.

On the other hand, from the boundary perspective, the intrinsic and extrinsic topological surface states are distinguishable through the attachment of topological states on the boundary of a crystal, called the surface decoration. Extrinsic topological surface states are constructed by pasting lower-dimensional topological superconductors (TSCs) to the boundary of the crystal faces. Removing the extrinsic topological surface states from all possible surface states results in the boundary classifying group $\mathcal{K}_a^{(n)}$, which denotes the intrinsic surface states of codimension $n$. The bulk-boundary correspondence manifests in the relationship between the subgroup series (\ref{eq:k-series}) and $\mathcal{K}_{\rm a}^{(n)}$,
\begin{align}
    \mathcal{K}_{\rm a}^{(n+1)} = K^{(n)}/K^{(n+1)}, \ \ n=0,1,2, \label{eq:bulk-boundary}
\end{align}
where $K^{(0)} \equiv K$. 

From the bulk classification, $\mathcal{K}_a^{(n)}$ are determined as in Table~\ref{tab:pairings}, where $\mathcal{K}_a^{(1)}=0$ for all pairing states since there is no first-order topological phase in class D. We can obtain the same results from the boundary classification in Appendix~\ref{app:boundary}. In addition, we identify topological invariants relevant to HOTPs in Sec.~\ref{sec:topo_inv} and predict the possible configuration of surface states as illustrated in Figure~\ref{Fig:Boundary-states}. The comprehensive classification of intrinsic surface states for possible pairing symmetries of UTe$_2$ clarifies that 25 out of 36 pairing states can potentially realize HOTPs. It should be noted that the classification results differ from those obtained under an order-two symmetry~\cite{Luka011012} due to multiple order-two symmetries.

\subsection{Topological invariants}
\label{sec:topo_inv}
To see the implication of the topological classification, we consider the topological invariants defined in the Bogoliubov-de Gennes (BdG) Hamiltonian,
\begin{align}
H &= \frac{1}{2}\sum_{\bm{k}} \bm{c}^{\dagger}_{\bm{k}} H(\vv k) \bm{c}_{\bm{k}}, \\
H(\vv k) &=  \left[ 
\begin{array}{cc}
  \epsilon(\vv k) - \mu & \Delta(\vv k) 
  \\
  \Delta^{\dagger}(\vv k) & -\epsilon^{\text{T}}(-\vv k) + \mu
\end{array} 
\right],
\label{BdG22}
\end{align}
where $\bm{c}_{\bm{k}} = [c_{\bm{k}},c^{\dagger}_{-\bm{k}}]^{\text{T}}$ and $c_{\bm{k}}$ ($c_{\bm{k}}^{\dagger}$) is the annihilation (creation) operator of the electron with momentum $\bm{k}$, which implicitly includes the indices for the spin, orbital, and sublattice degrees of freedom.
$\epsilon(\bm{k})$, $\Delta(\bm{k})$, and $\mu$ are the normal-state Hamiltonian, pair potential, and chemical potential, respectively. The BdG Hamiltonian satisfies the particle-hole (PH) symmetry as
\begin{align}
C H(\vv k) C^{-1} = -H(-\vv k), \ C=\tau_x K,
\end{align}
where the Pauli matrices $\tau_i$ ($i=x,y,z$) act on the Nambu space and $K$ is the complex conjugate operation. We assume that the normal-state Hamiltonian preserves the TR and crystalline symmetries,
\begin{align}
    &T \epsilon(\vv k) T^{-1} = \epsilon(-\vv k), \ T=is_y K, \\
    &D(g) \epsilon(\vv k) D(g)^{-1} = \epsilon(g\bm{k}), \ g \in mmm,
\end{align}
where the Pauli matrices $s_i$ ($i=x,y,z$) act on the spin space, $D(g)$ is the spinful unitary representation of $g$, and $g\bm{k}$ means the $O(3)$ transformation of $\bm{k}$ in terms of $g$. We fix the phase of $D(g)$ as $[T,D(g)]=0$.

 On the other hand, we assume that the TR symmetry is spontaneously broken in the pair potential, which is divided into TR-preserving and TR-breaking terms as
 \begin{align}
     \Delta_{ab}(\bm{k}) = [\Delta_{a}(\bm{k}) + i \Delta_{b}(\bm{k})] (i s_y), \label{eq:mixed-pair}
 \end{align}
where $a,b$ label the IRs of $mmm$, and each pair potential satisfies
\begin{align}
     &T \Delta_{a}(\bm{k}) T^{-1} = \Delta_{a}(-\bm{k}), \\
     &D(g) \Delta_{a}(\bm{k}) D^{\text{T}} (g)= \eta_{g,a} \Delta_{a}(g\bm{k}) \ \ g \in mmm,
\end{align}
where $\eta_{g,a} = \pm 1 $ encodes the information about IRs of $mmm$ in Table~\ref{tab:ir-mmm}. For instance, $\Delta_{A_u}$ satisfies $\eta_{g,A_u}=-1$ for $g=I, M_x,M_y,M_z$, otherwise $\eta_{g,A_u}=1$. Equation (\ref{eq:mixed-pair}) is invariant under the symmetry operations of $H \subseteq mmm$. Hence, $\Delta_{ab}$ satisfies 
\begin{align}
    D(g)\Delta_{ab} D^{\text{T}} (g) &= \eta_{g,a} \Delta_{ab}(g\bm{k}) \ \  g \in H. 
\end{align}
In the Nambu space, the crystalline symmetry operations are represented as
\begin{align}
    &\tilde{D}(g) H(\vv k) \tilde{D}^{-1}(g) = H(g\vv k), \ g \in H,  \label{eq:nambug} 
\end{align}
where $\tilde{D}(g) = \text{diag}[ D(g),\eta_{g,a} D^{\ast}(g) ]$ that satisfies 
\begin{align}
    \tilde{D}(g)C = \eta_{g,a} C \tilde{D}(g), \label{eq:commu_CD}
\end{align}
Here, $a$ is the IR of the TR-preserving part of $\Delta_{ab}$, and $\eta_{g,a}$ determines the commutation relation between $C$ and $\tilde{D}(g)$.

Similarly, when we fix the gauge as in Eq.~(\ref{eq:mixed-pair}), $\Delta_{ab}$ is transformed under $Th \in T(G-H)$ as
\begin{align}
    TD(h)\Delta_{ab} [TD (h)]^{\text{T}} &= \eta_{h,a} \Delta_{ab}(-h\bm{k}) 
\end{align}
Thus, the corresponding operation in the Nambu space is represented as
\begin{align}
    &\tilde{D}(Th) H(\vv k) \tilde{D}^{-1}(Th) = H(-h\vv k), \ h \in T(G-H), \label{eq:nambug}  
\end{align}
with $\tilde{D}(Th) = \text{diag}[TD(h), \eta_{h,a} T^{\ast}D^{\ast}(h)]$.
Hereafter, we omit the subscript of $\eta_{g,a}$ as $\eta_{g}$, unless otherwise specified.

\subsubsection{Possible topological invariants}\label{sec:possible}
The intrinsic HOTPs are characterized by bulk topological invariants.  Since $mmm$ is a symmorphic group and comprises the order-two symmetry operations that satisfy $D^2(I)=-D^2(2_i) = -D^2(m_i) = 1$ ($i=x,y,z$), possible topological invariants are the Chern number $\text{Ch}_1 \in \mathbb{Z}$, the mirror Chern number $\text{Ch}_1[g] \in \mathbb{Z}$ ($g=m_i$; $i=x,y,z$), the 1D magnetic winding number $W[h] \in \mathbb{Z}$ ($Th=T2_i,Tm_i$; $i=x,y,z$), the 1D crystalline $\mathbb{Z}_2$ topological invariants $\nu[g]_{\pm} \in \mathbb{Z}_2$ ($g=2_i,m_i$; $i=x,y,z$), the inversion symmetry indicator $\kappa[I] \in \mathbb{Z}_8$, and  the inversion symmetry indicator on mirror planes $\kappa[I]^{m_i}_{\pm} \in \mathbb{Z}_4$. Here, the subscript of $\nu[g]_{\pm} $ and $\kappa[I]^{m_i}_{\pm}$ labels the eigenspace of $\tilde{D}(g)=\pm i$ and $\tilde{D}(m_i)=\pm i$, respectively. The explicit definitions of these topological invariants are shown in Appendix~\ref{app:invariant}. The inversion symmetry indicators $\kappa[I]$ and $\kappa[I]^{m_i}_{\pm}$ are further decomposed into the $\mathbb{Z}_2$ indices~\cite{Anastasiia013064},
\begin{align} 
&\kappa[I] = \nu_1[I] + 2 \nu_2[I] + 4 \nu_3[I], \\
&\kappa[I]^{m_i}_{\pm} =\nu_1[I]^{m_i}_{\pm} + 2 \nu_2[I]^{m_i}_{\pm},
\end{align} 
where $\nu_1[I]=1$ indicates the existence of point nodes~\cite{Ono115150}. On the other hand, $\nu_3[I]=1$ and $\nu_2[I]=1$ ($\nu_2[I]^{m_i}_{\pm}=1$ and $\nu_1[I]^{m_i}_{\pm}=1$) indicate the existence of Majorana corner and hinge states on the inversion-symmetric geometry (in the mirror plane), respectively.  Note that some of them are related to each other. For instance, $\nu_1[I]=1$ implies $\text{Ch}_1 \neq 0$, and $\nu_1[I]^{m_i}_{+} = \nu_1[I]^{m_i}_{-}=1$ implies $\text{Ch}_1[m_i] \neq 0$.

Possible topological invariants for each IR of the magnetic point groups are listed in Table~\ref{tab:topology}, characterizing nodal superconducting phases, extrinsic HOTPs, and intrinsic HOTPs.

\begin{table}
	\caption{Possible topological invariants are shown for the magnetic point groups $mmm$, $m'm'm$, $mmm'$, and $m'm'm'$, where $\text{Ch}_1$ is the Chern number,  $\text{Ch}_1[m_i]$ the mirror Chern number, $W[h_i]$ the 1D magnetic winding number, $\nu[g_i]$ the 1D crystalline $\mathbb{Z}_2$ topological invariant, $\kappa[I]$ the inversion symmetry indicator, $\kappa[I]^{m_i}_{\pm}$ the inversion symmetry indicator on the $m_i$ mirror plane, where “$0$” indicates the absence of topological invariants. Topological invariants for other magnetic point groups are obtained by the permutation of $x,y,z$. 
    } \label{tab:topology}
		\begin{tabular}{cccc}
            \hline \hline
		  $M$  & IR & Pairings &  Topological invariants 
			\\
			\hline    
            $mmm$  & $A_g$ & $A_g + i A_g $ & $0$ 
            \\
            $mmm$  & $B_{1g}$ & $B_{1g} + i B_{1g}$&$\text{Ch}_1[m_{i}], \nu[2_i], \nu[m_i] $ ($i=x,y$) 
            \\
            $mmm$  & $B_{2g}$ & $B_{2g} + i B_{2g} $&$\text{Ch}_1[m_{i}], \nu[2_i], \nu[m_i]$ ($i=z,x$) 
            \\
            $mmm$ & $B_{3g}$ & $B_{3g} + i B_{3g}$&$\text{Ch}_1[m_{i}], \nu[2_i], \nu[m_i] $ ($i=y,z$) 
            \\
            $mmm$  & $A_u$ & $A_u + i A_u$ & $\text{Ch}_1[m_{i}], \kappa [I]^{m_i} $ ($i=x,y,z$) 
            \\
            $mmm$ & $B_{1u}$ & $B_{1u} + i B_{1u}$ &$\text{Ch}_1[m_{z}], \nu[m_z],\kappa [I]^{m_z},$ 
            \\
            &&& $\nu[2_x],\nu[2_y]$ 
            \\ 
            $mmm$   & $B_{2u}$ & $B_{2u} + i B_{2u}$ &$\text{Ch}_1[m_{y}], \nu[m_y], \kappa [I]^{m_y},$ 
            \\
            &&& $\nu[2_z],\nu[2_x]$ 
            \\
            $mmm$  & $B_{3u}$ & $B_{3u} + i B_{3u}$ &$\text{Ch}_1[m_{x}], \nu[m_x],  \kappa [I]^{m_x},$ 
            \\
            &&& $\nu[2_y],\nu[2_z]$ 
            \\
            $m'm'm$ & $A_g$ & $A_g + i B_{1g}$ & $\text{Ch}_1$  
            \\
            $m'm'm$  & $B_g$ & $B_{2g} + i B_{3g}$ &$\text{Ch}_1,\text{Ch}_1[m_{z}],\nu[2_z]_{\pm},\nu[m_z]_{\pm}$ 
            \\
            $m'm'm$  & $A_u$ & $A_u + i B_{1u}$ &$\text{Ch}_1, \text{Ch}_1[m_{z}],\nu[m_z]_{\pm} ,\kappa [I]^{m_z}_{\pm},$ 
            \\
            &&& $W[2_i],W[m_i]$ ($i=x,y$) 
            \\
            $m'm'm$  & $B_u$ & $B_{2u} + i B_{3u}$ & $\text{Ch}_1,W[m_x],W[m_y], \nu[2_z]_{\pm}, \kappa[I]$ 
            \\
            $mmm'$  & $A_1$ & $A_g + i B_{1u}$ &  $0$  
            \\
            $mmm'$  & $A_2$  & $B_{1g} + i A_u$ &$\text{Ch}_1[m_{i}], W[2_i], \nu[m_i]$ ($i=x,y$) 
            \\
            $mmm'$&  $B_1$& $B_{2g} + i B_{3u}$ & $\text{Ch}_1[m_{x}], W[m_z],\nu[2_z],\nu[m_x]$ 
            \\ 
            $mmm'$  & $B_2$ & $B_{3g} + i B_{2u}$ & $\text{Ch}_1[m_{y}], W[m_z],\nu[2_z],\nu[m_y]$ 
            \\
            $m'm'm'$  & $A$  & $A_g + i A_u$ & $0$ 
            \\
            $m'm'm'$  & $B_1$  & $B_{1g} + i B_{1u}$ & $\nu[2_i], W[m_i]$ ($i=x,y$)
            \\
            $m'm'm'$  & $B_2$  & $B_{2g} + i B_{2u}$ & $\nu[2_i], W[m_i]$ ($i=z,x$) 
            \\
            $m'm'm'$  & $B_3$  & $B_{3g} + i B_{3u}$ & $\nu[2_i], W[m_i]$ ($i=y,z$)
		  \\
            \hline \hline
 \end{tabular}	
\end{table}

\subsubsection{Nodal superconducting phases}
\label{sec:node}
 Topological invariants defined in any subspace of 3D Brillouin zone (BZ) characterize nodal superconducting phases. For instance, the Chern number is defined in any closed 2D subspaces, characterizing point nodes. Thus, $\text{Ch}_1 \neq 0$ implies that the BdG Hamiltonian is in the Weyl superconducting phase with the surface Majorana arc states~\cite{Tobias054504}. Similarly, the 1D magnetic winding number in terms of mirror-reflection symmetry $W[m_i]$ is defined in any closed 1D subspaces in the mirror plane~\cite{kobayashi14}. Thus, $W[m_i] \neq 0$ implies that the BdG Hamiltonian is in Dirac or Weyl superconducting phases with the surface Majorana flat band states on the mirror plane~\cite{Yang046401}. From the symmetry constraint, the point nodes are absent in case (i), and the Chern number is zero in case (iii). It is worth noting that nodal HOTPs are possible when $W[m_i] \neq 0$~\cite{ Simon2022,Henrik054521}.

As discussed in Sec.~\ref{sec:wca}, superconducting nodes also arise from the obstruction to forming the Cooper pair on the Fermi surface due to crystalline symmetry constraints in high-symmetry subspaces~\cite{Sigrist239}.

\subsubsection{Higher-order topological phases}

The topological invariants relevant to HOTPs can be identified by subtracting the redundancy and the topological invariants associated with nodal superconducting phases from the set of possible topological invariants. The topological invariants for intrinsic HOTPs are obtained by further subtracting those for extrinsic HOTPs through the surface decoration. For example, $\nu[2_i]$ in $M=m'm'm'$ corresponds to an extrinsic third-order topological phase so that a $2_i$ symmetry-protected corner state is removable through the surface decoration (see Appendix~\ref{app:boundary}).  
The remaining topological invariants correspond to intrinsic HOTPs, which are consistent with the topological classification presented in Table~\ref{tab:pairings}.

Topological invariants provide insights into the configuration of topological surface states. Intrinsic second-order topological phases are characterized by the mirror Chern number $\text{Ch}_1[m_i] \in \mathbb{Z}$, which corresponds to a helical Majorana hinge state in the mirror plane. In contrast, intrinsic third-order topological phases are classified by $\mathbb{Z}_2$ and $\mathbb{Z}$, which are linked to the inversion symmetry indicator $\nu_3 [I]$ ($\nu_2[I]^{m_i}_{\pm}$ in the mirror plane) and the 1D magnetic winding number $W[2_i]$, respectively. The inversion symmetry indicator predicts a pair of Majorana corner states at antipodal points in inversion-symmetric geometry, while the 1D magnetic winding number predicts pairs of double Majorana corner states at the rotation-symmetric corners of the crystal. 

In addition, these topological invariants are linked to each other. We examine topological invariants of minimal models that form bases of the classifying groups. The calculations are relegated to Appendix~\ref{app:generator}.  Figure~\ref{Fig:Boundary-states} displays the possible configuration of topological surface states characterized by a set of the topological invariants. For instance, we have three minimal sets of topological invariants: $(-1,-1,1,1)$, $(-1,1,-1,1)$, and $(1,-1,-1,1) \in (\text{Ch}_1[m_x],\text{Ch}_1[m_y],\text{Ch}_1[m_z],\nu_2[I]^{m_z}_+)$ for the $A_{u}+iA_u$ pairing states. These sets all correspond to the configuration with three helical Majorana hinge states on the three orthogonal mirror planes. The relationship reveals an intricate interplay between Majorana hinge and corner states, resulting in the coexistence of multiple helical Majorana hinge states in different mirror planes and Majorana corner states at the rotation axis.

In particular, for the $A_u + iB_{ju}$ and $B_{jg} + i A_u$ pairings ($j=1,2,3$), the 1D magnetic winding number $W[2_i]$ and the mirror Chern number $\text{Ch}_1[m_j]$ ($i\neq j$) are related by
\begin{align}
 (-1)^{\frac{W[2_i]}{2}} = (-1)^{\text{Ch}_1[m_j]},   \label{eq:mirror-rot}
\end{align}
where $W[2_i]$ must be an even integer for a fully gapped superconductor (see Appendix~\ref{app:generator}). Equation~(\ref{eq:mirror-rot}) implies that the 1D magnetic winding number features the second-order boundary state when $\frac{W[2_i]}{2}$ is an odd integer. This property generalizes the case involving $T2_i$ symmetry, with the topological classification given by $\mathcal{K}_a^{(2)} = \mathbb{Z}_2$ and $\mathcal{K}_a^{(3)} = 2\mathbb{Z}$~\cite{Luka011012}. This result indicates that the odd values of the 1D magnetic winding number are associated with second-order boundary states, while even values correspond to third-order boundary states. The classification of the second-order topological phase changes from $\mathbb{Z}_2$ to $\mathbb{Z}$ due to mirror-reflection symmetry.

\section{Weak coupling case}
\label{sec:wca}
In this section, we examine possible HOTPs in the weak coupling regime, where we assume that the energy scale of the pair potential is much smaller than that of the band hybridization, which imply that inter-band pairing is negligible. This assumption holds for weak-coupling superconductors. Under this assumption, a relationship emerges between the Fermi surface topology, superconducting nodes, and topological invariants, which imposes an additional constraint on pairing symmetry.   

Since the normal-state Hamiltonian preserves $T$ and $I \in mmm$ symmetries, the energy band is doubly degenerate due to $TI$ symmetry, which is referred to as the pseudospin degrees of freedom.  When inter-band pairing is neglected, a Cooper pair forms between electrons in an energy band with pseudospins, described by 
\begin{align}
    \Delta(\bm{k}) = [\psi(\bm{k}) + \bm{d}(\bm{k}) \cdot \bm{s}] i s_y,
\end{align}
where the Pauli matrices $s_i$ ($i=x,y,z$) describe the pseudospin degrees of freedom. The terms $\psi$ and $\bm{d}$ represent the components of pseudospin singlet and triplet pairings, respectively.
Under $mmm$ symmetry, the symmetry-adopted forms of pair potentials are given by, up to the quadratic terms of $k$,
\begin{subequations}
\label{eq:psinglet}
\begin{align}
&\psi_{A_g}(\bm{k}) = \rho_0+\rho_{x^2} k^2_x+\rho_{y^2} k^2_y+\rho_{z^2} k^2_z, \\ 
&\psi_{B_{1g}}(\bm{k}) = \rho_{xy} k_xk_y, \\ 
&\psi_{B_{2g}}(\bm{k}) = \rho_{zx} k_zk_x, \\
&\psi_{B_{3g}}(\bm{k}) = \rho_{yz} k_yk_z,  
\end{align}
\end{subequations}
for the pseudospin-singlet pairs, and
\begin{subequations}
\label{eq:ptriplet}
\begin{align}
&\bm{d}_{A_u}(\bm{k}) = (\rho_{x} k_x, \rho_{y} k_y, \rho_{z} k_z), \\ 
&\bm{d}_{B_{1u}}(\bm{k}) = (\rho_{y}k_y,\rho_{x}k_x,0), \\ 
&\bm{d}_{B_{2u}}(\bm{k}) = (\rho_{z}k_z,0,\rho_{x}k_x),  \\
&\bm{d}_{B_{3u}}(\bm{k}) = (0,\rho_{z}k_z,\rho_{y}k_y), 
\end{align}
\end{subequations}
for the pseudospin-triplet pairs, where $\rho_{i}$ are real coefficients. TRSB pairing states are constructed from the combination of Eqs.~(\ref{eq:psinglet}) and (\ref{eq:ptriplet}) as $\Delta_{ab} = \Delta_a + i \Delta_b$ with $a,b $ being the IRs of $mmm$. 

\subsection{Superconducting node structures}
\label{sec:node}
The crystalline-symmetry-protected superconducting nodes arise when $\Delta_{ab}=0$ in the high symmetric subspace such as the mirror planes and the rotation axes~\cite{Sigrist239}. For instance, the pair potential of $B_{2g} + i B_{3g}$ pairing is given by 
\begin{align}
    \Delta_{B_{2g} B_{3g}}(\bm{k}) &= [\psi_{B_{2g}}(\bm{k}) + i \psi_{B_{3g}}(\bm{k})](is_y) \nonumber \\
&= (\rho_{zx} k_zk_x+ i \rho_{yz} k_yk_z)(is_y).
\end{align}
This function vanishes at the $k_z=0$ plane and the $k_x=k_y=0$ line, leading to line and point nodes if the Fermi surface intersects with the high symmetric subspaces. 

In addition, when the Chern number $\text{Ch}_1$ or the 1D magnetic winding number $W[m_i]$ is nonzero, a point node apppars at an arbitrary $k$-point. One example is the $B_{2u}+iB_{3u}$ pairing state, whose pair potential is expressed as
\begin{align}
&\Delta_{B_{2u} B_{3u}}(\bm{k}) \nonumber \\
&= [\bm{d}_{B_{2u}}(\bm{k}) + i\bm{d}_{B_{3u}}(\bm{k})] \cdot \bm{s} (i s_y) \nonumber \\
&= [(\rho_x k_x + i\rho'_y k_y) s_z + k_z(\rho_z s_x + i\rho'_z s_y) ](is_y),
\label{chiral-fullgap}
\end{align}
where $\rho_i$ and $\rho'_i$ are real coefficients for $B_{2u}$ and $B_{3u}$, respectively. When $\rho_z=\rho_z'=0$, this pairing state realizes the $p_x + ip_y$ chiral pairing with point nodes located on the $k_x=k_y=0$ line. As $\rho_z$ and $\rho_z'$ vary, the point nodes move away from this line, with two pairs of point nodes shifting across the Fermi surface~\cite{Kozii2016,Maeno2023}. The positions of point nodes is given by 
\begin{subequations}
    \begin{align}
    &k_x k_y =0, \\
    &(\rho_x k_x)^2 + (\rho_z k_z)^2 = (\rho_y' k_y)^2 + (\rho_z' k_z )^2,
\end{align}
\end{subequations}
on the Fermi surface. When $|\rho_z| \neq |\rho'_z|$ and $|\rho_x|,|\rho'_y|<|\rho_z|,|\rho'_z|$, the point nodes pairwise annihilate, resulting in a fully gapped phase. The nodal superconducting phase corresponds to the Weyl superconducting phase discussed in Refs.~\cite{Shishidou104504,Choi2024}. Similarly, point nodes also appear at arbitrary $k$-points in the cases of $A_g + i B_{1g}$ and $A_u + iB_{1u}$ due to the presence of the Chern number and the 1D magnetic winding number.
 The superconducting node structures are summarized in Table~\ref{tab:node}.

\begin{table}
	\caption{ Superconducting node structures are presented for the magnetic point groups $mmm$, $m'm'm$, $mmm'$, and $m'm'm'$, under the weak coupling assumption. The fourth column indicates the zeros of the pair potentials, where nodes form at intersections with the Fermi surfaces. The fifth column classifies the gap structures as full gap (F), line node (L), or point node (P), where “arbitrary $k$-points" refers to point nodes that are not constrained to high symmetry lines but instead move across the Fermi surface. Their stability is ensured by $\text{Ch}_1$ or $W[m_i]$. For the $A_g + iB_{1g}$ pairing state, point nodes arise only when the zeros of $\psi_{A_g} (\bm{k})$ intersect the Fermi surface, which requires a sign change in the $A_g$ pairing component. Node structures for other magnetic point groups can be obtained by permuting the $x,y,z$ coordinates. } \label{tab:node}
		\begin{tabular}{ccccc}
            \hline \hline
		  $M$  & IR & Pairings &  Zeros of $\Delta$ & Gap
			\\
			\hline    
            $mmm$  & $A_g$ & $A_g + i A_g $ & -- & F
            \\
            $mmm$  & $B_{1g}$ & $B_{1g} + i B_{1g}$& $k_x=0$ or $k_y=0$ & L
            \\
            $mmm$  & $B_{2g}$ & $B_{2g} + i B_{2g} $ & $k_z=0$ or $k_x=0$ & L
            \\
            $mmm$ & $B_{3g}$ & $B_{3g} + i B_{3g}$ & $k_y=0$ or $k_z=0$ & L
            \\
            $mmm$  & $A_u$ & $A_u + i A_u$ & -- & F
            \\
            $mmm$  & $B_{1u}$ & $B_{1u} + i B_{1u}$ &$k_x=k_y=0$ & P
            \\ 
            $mmm$  & $B_{2u}$ & $B_{2u} + i B_{2u}$  &$k_z=k_x=0$ & P
            \\
            $mmm$  & $B_{3u}$ & $B_{3u} + i B_{3u}$ &$k_y=k_z=0$ & P
            \\
            $m'm'm$ & $A_g$ & $A_g + i B_{1g}$ & arbitrary $k$-points & P
            \\
            $m'm'm$  & $B_g$ & $B_{2g} + i B_{3g}$ & $k_z=0$ or $k_x=k_y=0$ & L,P
            \\
             $m'm'm$  & $A_u$ & $A_u + i B_{1u}$ & arbitrary $k$-points & P
            \\
            $m'm'm$  & $B_u$ & $B_{2u} + i B_{3u}$  & arbitrary $k$-points & P
            \\
            $mmm'$  & $A_1$ & $A_g + i B_{1u}$  &  --  & F
            \\
            $mmm'$  & $A_2$  & $B_{1g} + i A_u$ & -- & F
            \\
            $mmm'$&  $B_1$& $B_{2g} + i B_{3u}$ & $k_y=k_z=0$ & P
            \\ 
            $mmm'$  & $B_2$ & $B_{3g} + i B_{2u}$ &$k_z=k_x=0$ & P
            \\
            $m'm'm'$  & $A$  & $A_g + i A_u$  & -- & F
            \\
            $m'm'm'$  & $B_1$  & $B_{1g} + i B_{1u}$ & $k_x=k_y=0$ & P
            \\
            $m'm'm'$  & $B_2$  & $B_{2g} + i B_{2u}$  & $k_z=k_x=0$ & P
            \\
            $m'm'm'$ & $B_3$  & $B_{3g} + i B_{3u}$ & $k_y=k_z=0$ & P
		  \\
            \hline \hline
 \end{tabular}	
\end{table}

\subsection{Weak-coupling conditions for intrinsic HOTPs}
\label{sec:FS_formula}
Under the assumption, we can describe the 1D magnetic winding numbers and inversion symmetry indicators using Fermi surface formulae~\cite{Sato214526, Fu097001, Sato220504, Sato224511,Anastasiia013064}, which allow us to calculate the topological invariants from information about Fermi surface in the normal-state Hamiltonian. Since the 1D magnetic winding numbers and inversion symmetry indicators feature third-order boundary states and a part of second-order boundary states, we can relate these topological phases to the Fermi surface topology. 
From Table \ref{tab:node}, some of pairing symmetries have nodes on the high symmetric plans and lines. Thus, a candidate of intrinsic HOTPs is restricted to $A_u+iB_{ju}$, $B_{jg} + iA_u$, and $B_{ju} + i B_{ku}$ pairings ($j,k=1,2,3$; $j \neq k$), where the $A_u+iB_{ju}$ and $B_{ju} + i B_{ku}$ pairings states are in a fully gapped phase when all Weyl points are pair-annihilated. 
In the following, we discuss the Fermi surface formulae of the $A_u + iB_{1u}$, $B_{1g} + i A_u$ and $B_{2u} + i B_{3u}$ pairings, assuming 3D Fermi surfaces enclosing a TRIM, say, the $\Gamma$ point. The formulae are applied to the other pairing states in the same symmetry class.

\subsubsection{ $A_u + iB_{1u}$ pairing}
In this symmetry class, we have $\mathcal{K}_a^{(2)}=\mathbb{Z}$ and $\mathcal{K}_a^{(3)}=\mathbb{Z}^2$, characterized by $\text{Ch}_1[m_z]$ and $W[2_i]$ ($i=x,y$), respectively. 
The Fermi surface formula for the 1D magnetic winding numbers is given by~\cite{Sato224511}
\begin{align}
    W[2_i] = \frac{1}{2}  \sum_{\bm{k}_{\rm F} } \, \text{sgn} [v (\bm{k}_{\rm F}) \delta (\bm{k}_{\rm F})], \label{eq:FSw}
\end{align}
where Fermi points are defined as $\det[h(\bm{k}_{\rm F})-\mu]=0$ in a $2_i$ symmetric line, $v (\bm{k}_{\rm F})$ and $\delta (\bm{k}_{\rm F})$ are the Fermi velocity and pair potential at the Fermi points, $\text{sgn} [f]$ means the sign of $f$, and the summation is taken over all Fermi points $\bm{k}_{\rm F}$. The inversion symmetry leads to $v (-\bm{k}_{\rm F}) = - v (\bm{k}_{\rm F})$ and $\delta_{A_u B_{1u}}(-\bm{k}_{\rm F}) = -\delta_{A_u B_{1u}}(\bm{k}_{\rm F})$ on the symmetric line. Thus, Eq.~(\ref{eq:FSw}) is further simplified to
 \begin{align}
     &W[2_i] = \sum_{\bm{k}_{\rm F} >0 }\, \text{sgn} [v (\bm{k}_{\rm F})\delta_{A_u B_{1u}} (\bm{k}_{\rm F})], \label{eq:FSw2}
 \end{align}
 which depends on the number of Fermi surfaces that cross the symmetric line and the sign of pair potentials on each Fermi surface. 
Combining Eq.~(\ref{eq:FSw2}) with Eq.~(\ref{eq:mirror-rot}), we obtain the Fermi surface formula for the mirror Chern number as
\begin{align}
    (-1)^{\text{Ch}_1[m_z]} &= (-1)^{ \frac{1}{2}\sum_{\bm{k}_{\rm F} >0 }\, \text{sgn} [v (\bm{k}_{\rm F})\delta_{A_u B_{1u}}  (\bm{k}_{\rm F})]} \label{eq:FS-aub1u},
\end{align}
which enable us to identify the second-order topological phases from the Fermi surface topology.

\subsubsection{ $B_{1g} + i A_u$ pairing}
\label{sec:wca_b1g+iau}
From Table~\ref{tab:pairings}, the classifying groups are given by $\mathcal{K}_a^{(2)}=\mathbb{Z}^2$ and $\mathcal{K}_a^{(3)}=\mathbb{Z}^2$. The corresponding topological invariants are given by $\text{Ch}_1[m_i]$ and $W[2_i]$ ($i=x,y$). The mirror-reflection symmetry perpendicular to the $2_i$ rotation axis leads to $v (-\bm{k}_{\rm F}) = - v (\bm{k}_{\rm F})$ and  $\delta_{B_{1g} A_{u}}(-\bm{k}_{\rm F}) = -\delta_{B_{1g} A_{u}}(\bm{k}_{\rm F})$ on the $2_i$ symmetric line. Thus, we have the similar Fermi surface formulae as in Eq.~(\ref{eq:FSw2}) and (\ref{eq:FS-aub1u}) for $W[2_i]$ and Ch$_1 [m_j]$ ($i \neq j$).

\subsubsection{ $B_{2u} + i B_{3u}$ pairing}\label{main: B2u+iB3u}
In this class, the system exhibits only a third-order topological phase classified by $\mathcal{K}_a^{(3)} = \mathbb{Z}_2$ and characterized by $\nu_3 [I]$.
Under the assumption, the inversion symmetry indicator is expressed as~\cite{Anastasiia013064}
\begin{align} 
\kappa[I] = \sum_{\bm{k} \in \text{3D TRIM}}\left(n^{+}_{\bm{k},\text{N}}-n^{-}_{\bm{k}, \text{N}}\right) \quad \text{mod $8$}, \label{eq:inversion_indicator}
\end{align}
where $n^{\pm}_{\bm{k},\text{N}}$ denotes the number of occupied states for the normal-state Hamiltonian with an inversion eigenvalue of $\pm 1$ at a time-reversal-invariant momentum (TRIM). The third-order topological phase emerges when four Fermi surfaces enclose a TRIM. Table~\ref{tab:pairings} shows no intrinsic second-order topological phase due to the presence of the other order-two symmetries.

\section{Topological surface states in mixed-parity superconductors}
\label{sec:model}

We demonstrate topological surface states of HOTPs in a mixed-parity $A_u + iB_{1g}$ pairing state as an example. The classification and Fermi surface formulae predict second-order, hybrid-order, and third-order topological phases depending on the number of Fermi surfaces enclosing a TRIM (\#FS).        

\subsection{Model}

We construct a tight-binding model with $mmm$ symmetry, described as
\begin{align}
\epsilon(\bm{k}) &= c(\bm{k})+ t(\bm{k})\sigma_x + R_z \sin(k_z)\sigma_y \nonumber\\ 
&\quad \quad + [R_x \sin (k_y)s_x + R_y \sin (k_x)s_y]\sigma_z,  \label{eq:normal}
\end{align}
with
\begin{align}
c(\bm{k}) &= t_x \cos(k_x) + t_y \cos(k_y) + t_z \cos(k_z), \nonumber\\
t(\bm{k}) &= t_1 + t'_x \cos(k_x) + t'_y \cos(k_y) + t'_z \cos(k_z), \nonumber
\end{align}
 where $\bm{s}$ and $\bm{\sigma}$ denote the Pauli matrices in the spin and sublattice spaces. The $c(\bm{k})$ and $t(\bm{k})$ terms are the intra and inter sublattice nearest neighbor hopping terms, and the $R_x$, $R_y$, and $R_z$ terms are spin-orbit coupling terms. We choose the lattice constants to be $1$. The normal-state Hamiltonian preserves TR symmetry ($T=is_y K$) and $mmm$ symmetry, generated by $D(2_z)=-is_z$, $D(2_x) = -i s_x \sigma_x$, and $D(I) = \sigma_x$. 
 
 For the superconducting state, the $A_u + iB_{1g}$ pairing state is given by
\begin{align}
\Delta(\bm{k}) = [\Delta_{A_u}(\bm{k}) + i \Delta_{B_{1g}}(\bm{k})]is_y, \label{eq:pair-potential_au+ib1u}
\end{align}
with
\begin{align}
\Delta_{A_u}(\bm{k}) &= \Delta_0 s_z \sigma_y + \sum_{i=x,y,z} \Delta_i s_i \sin(k_i), \nonumber \\ 
\Delta_{B_{1g}}(\bm{k}) &= \sum_{i=x,y,z} \Delta'_i s_i \sigma_z \sin(k_i), \nonumber
\end{align}
where $\Delta_0$ is the on-site pair potential, $\Delta_i$ is the sublattice-independent nearest neighbor pair potentials, and $\Delta_i'$ is the sublattice-dependent nearest neighbor pair potentials. The pair potential (\ref{eq:pair-potential_au+ib1u}) belongs to the $A_2$ IR of $mmm'$ symmetry. Hence, the BdG Hamiltonian (\ref{BdG22}) is invariant under $mmm'$ symmetry, whose operators are  represented as 
\begin{align}
\tilde{D}(2_z)= -i s_z \tau_z, \, \tilde{D}(m_{x})= i s_x, \, \tilde{D}(m_{y})= i s_y \tau_z
\end{align}
for the unitary operators, and
\begin{align}
&\tilde{D}(TI) = -i s_y \sigma_x \tau_z K, \, \tilde{D}(T2_x)= -i s_z \sigma_x \tau_z K, \nonumber \\ 
&\tilde{D}(T2_y)= \sigma_x K, \, \tilde{D}(Tm_{z})= -is_x \sigma_x K, 
\end{align}
for the antiunitary operators.
Here, $\bm{\tau}$ is the Pauli matrix in the Nambu space. 

\subsection{Topological invariant and Fermi surface}

Based on Table \ref{tab:pairings} and the discussion in Section~\ref{sec:FS_formula}, the intrinsic HOTPs are characterized by ($\text{Ch}_1[m_{x}],\text{Ch}_1[m_{y}],W[2_x],W[2_y]$). The configuration of the topological surface states varies with changes in \#FS, where $\text{\#FS}$ is the number of 3D Fermi surfaces enclosing a TRIM. In the numerical calculation, the parameters are chosen to generate a 3D spherical Fermi surface enclosing the $\Gamma$ point. We consider two situations: $\text{\#FS}=2$ and $4$. The corresponding parameters are shown in Table~\ref{tab: parameter} \footnote{We choose that the magnitudes of $\Delta_{A_u}$ and $\Delta_{B_{1u}}$ are comparable. Thus, the situation differs from the HOTPs described in Ref. \cite{Henrik054521}, where $|\Delta_{B_{1u}}| \ll |\Delta_{A_u}|$ is assumed.}.
 Figure~\ref{Fig: numerical1} illustrates the Fermi surfaces for (a) \#FS $ = 2$ and (d) \#FS $ = 4$.
In these parameters, the topological invariants ($\text{Ch}_1[m_{x}],\text{Ch}_1[m_{y}],W[2_x],W[2_y]$) are numerically calculated as
\begin{align}
& (1,1,2,-2) \quad \text{if  \#FS=2}, \label{eq:top_FS=2} \\
&(-2,0,4,0) \quad \text{if  \#FS=4}, \label{eq:top_FS=4}
\end{align}
which agree with the results obtained from the Fermi surface formula. Equation (\ref{eq:top_FS=2}) reveals the presence of two Majorana hinge states in the $m_x$ and $m_y$ mirror planes. In contrast, Eq.~(\ref{eq:top_FS=4}) characterizes hybrid-order boundary states that exhibit a Majorana hinge state in the $m_x$ mirror plane and Majorana corner states on the $2_x$ rotation lines.  Figures \ref{Fig: numerical1} (b) and (e) provide schematic illustrations of these topological surface state configurations.

\subsection{Topological surface states}\label{sec: numerical}

\begin{figure*}[t] 
\centering
    \includegraphics[scale=0.051]{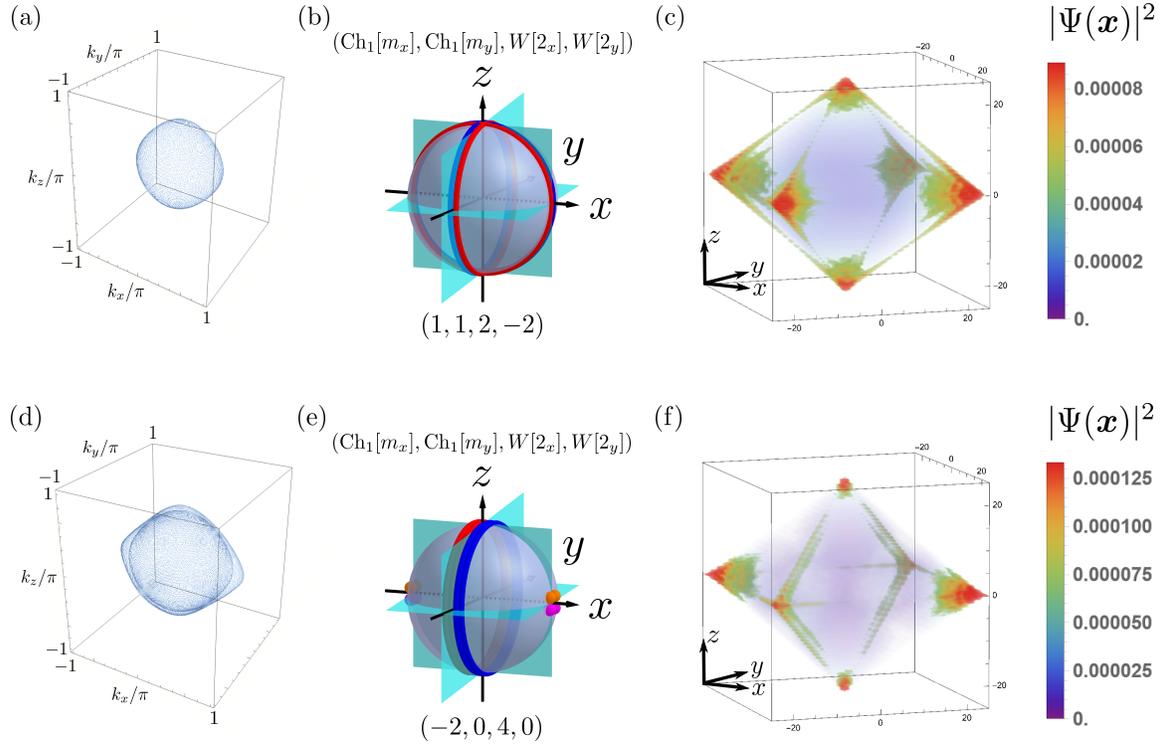}
    \caption{ 
    The 3D Fermi surface (a) [(d)], topological invariants (b) [(e)], and density of states for in-gap states (c) [(f)] are shown, where we use the parameters in Table~\ref{tab: parameter} when \#FS=2 [\#FS=4]. We plot the density of states of eigenstates satisfying $|\epsilon_n| \lesssim 0.25|E_{\text{gap}}|$, which is described as $|\Psi(\bm{x})|^2=(1/450)\sum^{450}_{n=1}|u_n(\bm{x})|^2 $ for \#FS=2 and $|\Psi(\bm{x})|^2=(1/150)\sum^{150}_{n=1}|u_n(\bm{x})|^2 $ for \#FS=4, where $u_n(\bm{x})$ and $\epsilon_n$ are eigenstates and eigenvalues of Eq. (\ref{eq:bdg_open}), and $E_{\text{gap}}$ is the minimum of the bulk energy gap.}
    \label{Fig: numerical1}
\end{figure*}

\begin{figure}[t] 
\centering
    \includegraphics[scale=0.3]{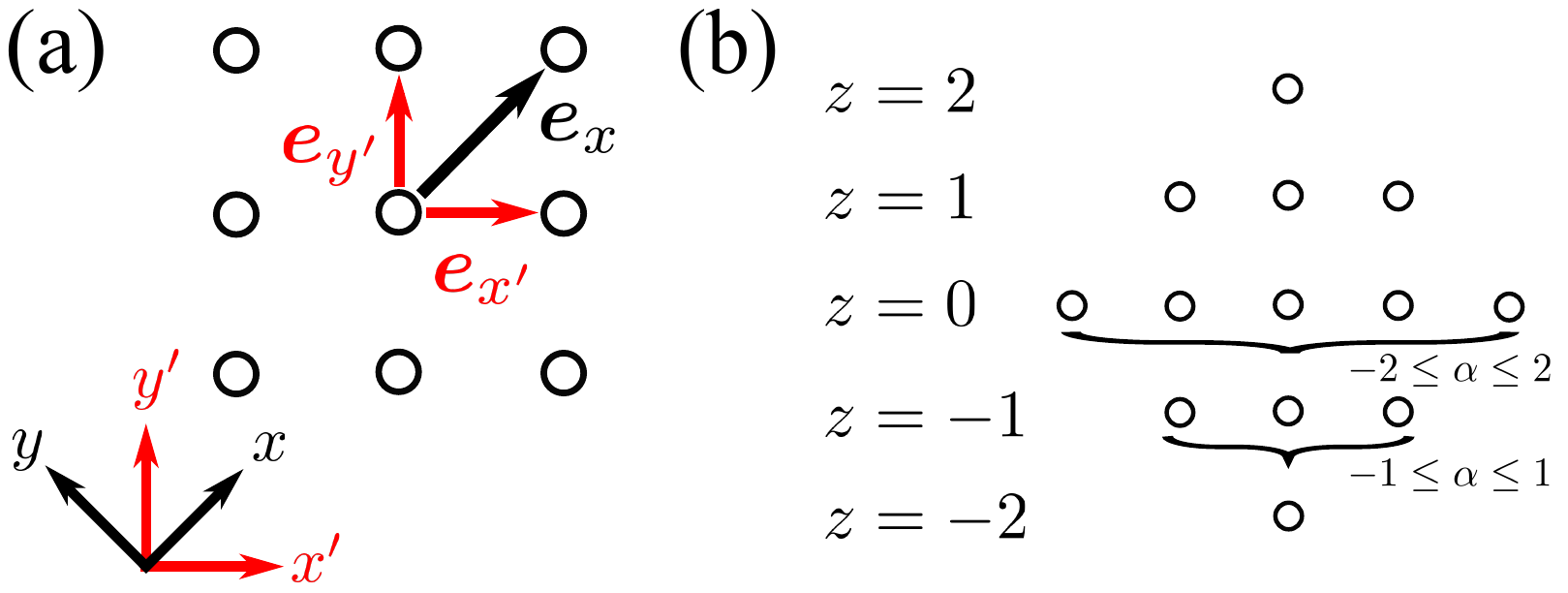}
    \caption{(a) The definition of unit vectors. (b) The lattice configuration of an octahedron with $L=5$, where $\alpha=x', y'$.}
    \label{Fig: numerical1}
\end{figure}

\begin{table}[t]
    \centering
    \caption{The parameters of the tight-binding model in Figure~\ref{Fig: numerical1}. We choose the parameters in such a way that the spherical Fermi surface encloses the $\Gamma$ point and satisfy either \#FS=2 or \#FS=4. For the numerical calculations, we fix $t_x=0.9,\ t_y=1.0, \ t_z=1.1, \ R_x=0.2, \ R_y=0.3, \ R_z = 0.4, \ \Delta_1 = 0.2$,  $\Delta'_x=0.05$, and  $\Delta'_z=0.1$.}
    \begin{tabular}{cccccccccccc}
    \hline \hline 
     \#FS & $\mu$ & $t_1$ & $t'_x$ & $t'_y$ & $t'_z$  & $\Delta_x$ & $\Delta_y$ & $\Delta_z$ & $\Delta'_y$ \\ \hline 
     $2$ & $3.5$ & $0.0$ &  $0.9$ & $1.0$ & $1.1$ & $0.1$ & $0.15$ & $0.05$ & $0.25$  
     \\
     $4$ & $2.0$ & $-5.25$ &  $3.15$ & $3.5$ & $3.85$ & $0.5$ & $0.05$ & $0.1$ & $0.4$ \\
     \hline \hline 
    \end{tabular}
    \label{tab: parameter}
\end{table}

To illustrate the boundary states, the BdG Hamiltonian with the normal-state Hamiltonian (\ref{eq:normal}) and the pair potential (\ref{eq:pair-potential_au+ib1u}) is numerically diagonalized in the real space. The crystal shape is set to be an octahedron, which is compatible with $mmm$ symmetry, i.e, the rotation axes and mirror planes coincide with the vertices and edges of the octahedron.

To build this configuration, we introduce a coordinate $(x',y',z) \equiv (x-y,x+y,z)$ and define the unit vectors as $\bm{e}_{x'} \equiv (\bm{e}_{x}-\bm{e}_{y})/2$, $ \bm{e}_{y'} \equiv (\bm{e}_{x}+\bm{e}_{y})/2$, and $\bm{e}_z$, as shown in Figure~\ref{Fig: numerical1} (a). 
The lattice site in the octahedron is then defined by $\bm{R} = x'\bm{e}_{x'} + y'\bm{e}_{y'} + z\bm{e}_z$, where the integers $x', y', z$ satisfy 
\begin{subequations}
\label{eq: real-space-coordinate}
\begin{align}
|z|-\frac{L-1}{2} \le &x' \le \frac{L-1}{2}-|z|, \\
|z|-\frac{L-1}{2} \le &y' \le \frac{L-1}{2}-|z|, \\
-\frac{L-1}{2} \le &z \le \frac{L-1}{2}. 
\end{align}
\end{subequations}
Here, $L$ is the size of the octahedron and must be an odd number.
For example, Figure~\ref{Fig: numerical1} (b) shows the lattice site in the octahedron when $L=5$. 
In this basis, the real space representation of the BdG Hamiltonian is in the form of
\begin{align}
\nonumber
H_{\text{open}} = \Big[&\sum_{\bm{R}}\bm{c}^{\dagger}_{x',y',z} H^{(0)} \bm{c}_{x',y',n_z} \\ \nonumber
+&\sum_{\bm{R}} \bm{c}^{\dagger}_{x',y',z} H^{(x)} \bm{c}_{x'+1,y'+1,z} \\ \nonumber
+&\sum_{\bm{R}} \bm{c}^{\dagger}_{x',y',z} H^{(y)} \bm{c}_{x'-1,y'+1,z} \\ \nonumber
+&\sum_{\bm{R}}\bm{c}^{\dagger}_{x',y',z} H^{(z)} \bm{c}_{x',y',z+1} \\ 
+&\sum_{\bm{R}} \bm{c}^{\dagger}_{x',y',z} H^{(z)} \bm{c}_{x',y',z+1}\Big] + \text{h.c.} \label{eq:bdg_open}
\end{align}
where $H^{(0)}$, $H^{(x)}$, $H^{(y)}$, and $H^{(z)}$ are the on-site term and the nearest-neighbor hopping terms in the $x$, $y$, and $z$ directions. The summation is taken over the lattice sites in the octahedron defined by Eq. (\ref{eq: real-space-coordinate}). 
In the numerical calculation, the size of the octahedron is set to $L=51$. Figure~\ref{Fig: numerical1} shows the density of states for the in-gap states for (c) \#FS = 2 and (f) \#FS = 4. As expected, the results show the double Majorna hinge states in Figure \ref{Fig: numerical1} (c), and the hybrid-order boundary state with  Majorana hinge and corner states in Figure~\ref{Fig: numerical1} (f). These boundary state configurations are consistent with the schematic illustrations in Figure~\ref{Fig: numerical1} (b) and (e).

\section{Conclusion}

We established a topological classification in 3D TRSB multi-component superconducting states of UTe$_2$ based on the symmetry classification of multi-component order parameters under the magnetic point groups and the K theoretical classification approach. The classification predicts the existence of a large variety of HOTPs even though the first-order topological phase is absent in 3D class D.  Therefore, UTe$_2$ is an intriguing playground to investigate not only the first-order TSC in class DIII but also higher-order TSCs in class D. In addition, we identified the configuration of the topological surface states based on the bulk topological invariants and the surface decoration, which display the coexistence of Majorana hinge and corner states in a complex way. For instance, the $A_{u}+iA_{u}$ pairing state hosts three Majorana hinge states on different mirror planes.

Assuming the weak-coupling superconductor, possible candidates of higher-order TSCs are restricted to $B_{ju} + i B_{ku}$, $A_{u} + i B_{j u}$, and $B_{jg} + iA_u$ ($j,k = 1,2,3$) due to inevitable superconducting nodes. For the candidates, we derived the Fermi surface formulae for HOTPs, which allow us to search for HOTPs from the information about the pair potential and the Fermi surface topology. In particular, the topological surface states of the mixed-parity $B_{1 g} + iA_u$ pairing states vary with change in the number of Fermi surfaces enclosing a TRIM, which exhibit the second-order, hybrid-order, and third-order topological surface states. Using a tight-binding model of TRSB superconductors with the $B_{j g} + iA_u$ pairing and numerically calculating a surface state under an open boundary condition that is compatible with the crystalline symmetry, we verified the second-order topological phase with double Majorana hinge states when \#FS $=2$, and it changes to the hybrid-order topological phase with Majorana hinge and corner states when \#FS $=4$. These anomalous surface states would be observed through electromagnetic response of Majorana quasiparticles~\cite{Yamazaki2024}, such as surface spin susceptibility ~\cite{Ohashi2024} and dynamic strain response~\cite{yamazakiL060505}. Proposing conclusive experiments for observing higher-order boundary states would be a desirable future task.

Finally, some remarks on the realization of HOTPs in superconducting UTe$_2$ are provided. First, the presence of a 3D Fermi surface is essential for achieving various configurations of topological surface states, including the coexistence of multiple Majorana hinge and corner states. Second, the $A_u$ paring state, which is supported by experiments on NMR Knight shift~\cite{matsumura063701} and thermal conductivity~\cite{Suetsugu2024} using high-quality crystal, plays a crucial role in both class DIII and class D. The pairing states $A_u + iA_u$, $A_u + i B_{ju}$, and $B_{jg} + i A_u$ each give rise to HOTPs with a hybrid-order topological boundary state. Third, a third-order topological phase arises in the $B_{ju} + iB_{ku}$ pairing states ($j \neq k$), which has previously been identified as the Weyl superconducting phase~\cite{Shishidou104504,Choi2024}. The HOTP arises when the number of Fermi surfaces is four, leading to a fully gapped phase with a pair of Majorana corner states protected by inversion symmetry. Fourth, multi-component order parameters that coexist with the $A_g$ IR are topologically trivial in the sense of 3D intrinsic HOTPs. Consequently, they only allow nodal superconducting phases or extrinsic HOTPs.  

\begin{acknowledgments}

S. K. was supported by JSPS KAKENHI (Grants No. JP22K03478 and No. JP24K00557). 
This work was supported by JST CREST (Grant No. JPMJCR19T2).

\end{acknowledgments}

\appendix

\section{IRs of $2mm$, $m2m$, and $mm2$.}

Table \ref{tab:ir-2mm} shows the definitions of IRs of $2mm$, $m2m$, and $mm2$ that we have adopted in this paper. The other point groups are consistent with the standard character table such as the Bilbao Crystallographic Server~\cite{Elcoro}.

\begin{table}
	\caption{Irreducible representations (IRs) of $2mm$, $m2m$, and $mm2$, where $E$ represents the identity operation, $2_i$ the twofold rotation around $i$ axis, and $m_{i}$ the mirror-reflection symmetry in terms of the plane normal to the $i$ axis.}
    \label{tab:ir-2mm}
		\begin{tabular}{ccccc}
            \hline 
            \multicolumn{5}{c}{$2mm$} 
            \\ \hline
            IR & $E$ & $2_x$ & $m_{y}$ & $m_{z}$ \\ \hline
            $A_1$ & $1$ & $1$ & $1$ & $1$  \\
            $A_{2}$ & $1$ & $1$ & $-1$ & $-1$ \\
            $B_{1}$ & $1$ & $-1$ & $-1$ & $1$  \\
            $B_{2}$ & $1$ & $-1$ & $1$ & $-1$ \\
            \hline \hline
            \end{tabular}
            \begin{tabular}{ccccccccc}
            \hline 
            \multicolumn{5}{c}{$m2m$} 
            \\ \hline
            IR & $E$ & $2_y$ & $m_{z}$ & $m_{x}$ \\ \hline
            $A_1$ & $1$ & $1$ & $1$ & $1$  \\
            $A_{2}$ & $1$ & $1$ & $-1$ & $-1$ \\
            $B_{1}$ & $1$ & $-1$ & $1$ & $-1$  \\
            $B_{2}$ & $1$ & $-1$ & $-1$ & $1$ \\
            \hline \hline
            \end{tabular}
            \begin{tabular}{ccccccccc}
            \hline 
            \multicolumn{5}{c}{$mm2$} 
            \\ \hline
            IR & $E$ & $2_z$ & $m_{y}$ & $m_{x}$ \\ \hline
            $A_1$ & $1$ & $1$ & $1$ & $1$  \\
            $A_{2}$ & $1$ & $1$ & $-1$ & $-1$ \\
            $B_{1}$ & $1$ & $-1$ & $1$ & $-1$  \\
            $B_{2}$ & $1$ & $-1$ & $-1$ & $1$ \\
            \hline \hline
            \end{tabular}
\end{table}

\section{Bulk classification}
\label{app:bulk}
In this Appendix, we explain how to calculate the subgroup series of classifying groups (\ref{eq:k-series}). The argument is based on the classification of 3D Dirac Hamiltonians~\cite{Shiozaki04A104,Luka202000090}. Dirac Hamiltonians describe a low-energy description of band structures close to a topological phase transition, which are represented as
\begin{align}
    H_{\rm D}(\bm{k}) = m \Gamma_0 + k_x \Gamma_1 + k_y \Gamma_2 + k_z \Gamma_3,  \label{eq:3dDirac}
\end{align}
where matrices $\Gamma_j$ ($j=0,\cdots,3$) are mutually anticommuting matrices and satisfy $\Gamma_j^2=1$. We assume that the 3D Dirac Hamiltonians satisfy PH symmetry $\hat{C}$ ($\hat{C}^2=1 $),
\begin{align}
    &\hat{C} H_{\rm D}(\bm{k}) \hat{C}^{-1} = -H_{\rm D}(-\bm{k}), \ \ \hat{C} i \hat{C}^{-1} = - i \nonumber \\
    & \Leftrightarrow \{\Gamma_0 ,\hat{C}\} = [\Gamma_k ,\hat{C}]=0, \ \ k=1,2,3. \label{eq:Dirac-PHS}
\end{align}
and magnetic point symmetry $\hat{g}$, 
\begin{align}
    &\hat{g}H_{\rm D}(\bm{k}) \hat{g}^{-1}  = H_{\rm D}(\phi_g O_g \bm{k}), \ \ \hat{g} i \hat{g}^{-1} = \phi_g i , \nonumber \\
     & \Leftrightarrow \hat{g} \Gamma_i \hat{g}^{-1} = \sum_j \phi_g [O_g^{-1}]_{ij} \Gamma_j, \ \ \hat{g} \Gamma_0 \hat{g}^{-1} = \Gamma_0 \label{eq:Dirac-g}
\end{align}
where $\phi_g=1(-1)$ for the (anti)unitary operator, and $O_g \in O(3)$ is a $3 \times 3$ real representation of $g$ that acts on $\bm{k}$ and $\bm{x}$. The operators $\hat{C}$ and $\hat{g}$ satisfy
\begin{align}
    \hat{g} \hat{C} = \eta_g \hat{C}  \hat{g}, 
\end{align}
where $\eta_g$ is determined from the IRs of pair potentials.

The classifying groups are calculated by using the Cornfeld-Chapman isomorphism~\cite{Chapman075105}, which gives a mappping between the Dirac Hamiltonians with a magnetic point group symmetry and those with an onsite symmetry. The mapping is constructed from an $SO(3)$ rotation operation about the $\bm{n}$ axis by $\theta$ and an antisymmetric spatial-inversion operation:
\begin{align}
   & \hat{g}_{\Gamma} H_{\rm D}(\bm{k}) \hat{g}_{\Gamma}^{-1} = H_{\rm D}(R_g^{-1} \bm{k}), \\ 
   & \hat{I}_{\Gamma}  H_{\rm D}(\bm{k}) \hat{I}_{\Gamma} ^{-1} = -H_{\rm D}(-\bm{k}),
\end{align}
where $\hat{g}_{\Gamma} = e^{\frac{\theta}{2} (n_1\Gamma_2 \Gamma_3 + n_2 \Gamma_3 \Gamma_1 + n_3\Gamma_1 \Gamma_2  )}$, $\hat{I}_{\Gamma} = \Gamma_1 \Gamma_2 \Gamma_3$, and $\bm{n} = (n_1,n_2,n_3)$ is a unit vector, $R_g$ is an $SO(3)$ part of $O_g \in O(3)$. The combination of $\hat{g}_{\Gamma}$ and $\hat{I}_{\Gamma}$ gives an element of $O(3)$, and  antiunitary operators are mapped to unitary operators by combining it with $\hat{C}$. Thus, the point group operators are mapped to on-site unitary operators by combining the original crystalline operations with $\hat{g}_{\Gamma}$, $\hat{I}_{\Gamma}$ and $\hat{C}$ as
\begin{subequations}
\label{eq:comm_K0}
   \begin{align}
    &\tilde{g}_{\rm o} = \begin{cases} e^{i \varphi_g} \hat{g}_{\Gamma} \hat{g}  &\text{if } s_g=\phi_g = 1\\ 
                                       e^{i \varphi_g} \hat{I}_{\Gamma} \hat{g}_{\Gamma} \hat{g}\hat{C}  &  \text{if } s_g=\phi_g = -1\end{cases} \\
    &\tilde{g}_{\rm c} =  \begin{cases}  e^{i \varphi_g}\hat{I}_{\Gamma} \hat{g}_{\Gamma} \hat{g}   &\text{if } s_g=-\phi_g = -1 \\
                            e^{i \varphi_g}  \hat{g}_{\Gamma} \hat{g}\hat{C}  &\text{if } s_g=-\phi_g = 1 \\ \end{cases}
\end{align} 
\end{subequations}
where $s_g \equiv \det(O_g)$ and $e^{i \varphi_g}$ is chosen to be $\tilde{g}^2_{\rm o(c)}=1$. These operations act on the 3D Dirac Hamiltonians as the unitary on-site symmetry,
\begin{align}
    &\tilde{g}_{\rm o} H_{\rm D}(\bm{k}) \tilde{g}_{\rm o}^{-1} = H_{\rm D}(\bm{k}), \\
    &\tilde{g}_{\rm c} H_{\rm D}(\bm{k}) \tilde{g}_{\rm c}^{-1} = -H_{\rm D}(\bm{k}),  
\end{align}
where the commutation relations between $\tilde{g}_{\rm o}$, $\tilde{g}_{\rm c}$, and $\hat{C}$ are different from the original operators. As a result, the calculation of the classifying group is reduced to the problem of determining the tenfold-way classes of the 3D Dirac Hamiltonian with the on-site symmetries, the classification of which is well known.

Given that the classifying group $K^{(0)}$ of the Dirac Hamiltonian (\ref{eq:3dDirac}) is determined, the next step is to determine the subgroups $K^{(n)}$ $(1\le n \le d)$, which are determined by calculating the classifying group $K_{O_n}$ of the 3D Dirac Hamiltonian with the $n$ mass terms $M_1, \cdots, M_n$, which is defined by 
\begin{align}
    H_{{\rm D}}^{O_n}(\bm{k},\bm{x}) = H_{\rm D} (\bm{k}) + \sum_{i=1}^n x_i M_i, \label{eq:Dirac_w_mass}
\end{align}
where $x_i$ are the real orthogonal coordinates, and $M_i$ are mutually anticommuting matrices satisfying $\{M_j ,H_{\rm D} (\bm{k})\}=0$. The mass terms $M_1, M_2, \cdots, M_n$ transform under real representations $O_n$ of the point group, each of which corresponds to $(n+1)$th order boundary signatures. The classifying group $K_{O_n}$ is  calculated using the Cornfeld-Chapman isomorphism as well. When $n=3$, it is sufficient to consider only a single representation of the point group, where $M_1$, $M_2$, and $M_3$ transform in the same way as the position vector. Hence, the real coordinate is chosen as $(x_1,x_2,x_3) = (x,y,z)$. Note that the classifying group $K_{O_3}$ classifies a bound state localized at the point group center, which is isomorphic to the classification of atomic limit phases~\cite{Shiozaki04A104,Luka202000090}. From the symmetry constraints~(\ref{eq:Dirac-PHS}) and (\ref{eq:Dirac-g}), $M_i$ satisfy
\begin{align}
       \{\hat{C},M_i\}=0, \ \ \hat{g} M_i \hat{g}^{-1} = \sum_j [O_g^{-1}]_{ij} M_j,
\end{align}
where $\bm{x} \to O_g \bm{x}$ for the antiunitary operation $\hat{g}$. 
Thus, the on-site operators are given by
\begin{subequations}
\label{eq:comm_Kn}
\begin{align}
    &\tilde{g}_{{\rm o}}^{O_3} = \begin{cases} e^{i \varphi_g} \tilde{g}_{M} \hat{g}_{\Gamma} \hat{g} &\text{if } s_g=\phi_g = 1, \\
      e^{i \varphi_g}  \tilde{I}_M\hat{I}_{\Gamma} \tilde{g}_{M} \hat{g}_{\Gamma} \hat{g}  &\text{if } s_g=-\phi_g = -1, \end{cases}\\
    &\tilde{g}_{{\rm c}}^{O_3} =\begin{cases}e^{i \varphi_g} \tilde{g}_{M} \hat{g}_{\Gamma} \hat{g}\hat{C}  &\text{if } s_g=-\phi_g = 1, \\
     e^{i \varphi_g} \tilde{I}_M\hat{I}_{\Gamma} \tilde{g}_{M} \hat{g}_{\Gamma} \hat{g}\hat{C}   &\text{if } s_g=\phi_g = -1, \end{cases} 
\end{align}
\end{subequations}
where $\tilde{g}_{M} = e^{\frac{\theta}{2} (n_1 M_2 M_3 + n_2 M_3 M_1 + n_3 M_1 M_2  )}$, $\hat{I}_{M} = M_1 M_2 M_3$, and the phase factors are added so that the square of the operators is $1$. These on-site unitary operators act on $H_{{\rm D}}^{O_3}(\bm{k},\bm{x})$ as
\begin{align}
    &\hat{g}_{\rm o}^{O_3} H_{\rm D}^{O_3}(\bm{k},\bm{x})(\hat{g}_{\rm o}^{O_3})^{-1} = H_{\rm D}^{O_3}(\bm{k},\bm{x}), \\
    &\hat{g}_{\rm c}^{O_3} H_{\rm D}^{O_3}(\bm{k},\bm{x})(\hat{g}_{\rm c}^{O_3})^{-1} = -H_{\rm D}^{O_3}(\bm{k},\bm{x}).
\end{align}
When $n=2$,  $K_{O_2}$ is determined in the similar way to $K_{O_3}$, where we need to consider all 2D real representations corresponding to third-order boundary signatures.  
When $n=1$, we need not to calculate $K_{O_1}$ because $K^{(1)}$ classify topological phases that exclude the first-order topological phases. That is, this subgroup is calculated from the kernel of the inclusion map $K^{(0)} \hookrightarrow K_{\rm TF}$, where $K_{\rm TF}$ is the tenfold-way classifying group without crystalline symmetries~\cite{Luka202000090}. In 3D TRSB superconductors, $K_{\rm TF} =0$, whereby leading to $K^{(0)} = K^{(1)}$.

Finally, the subgroup $K^{(n)} \subseteq K^{(0)}$ is generated by the images of the map $K_{O_n} \to K^{(0)}$, which is given by omitting the mass terms $M_1, M_2, \cdots, M_n$ in Eq.~(\ref{eq:Dirac_w_mass}).

\begin{table}
	\caption{ Bulk classification sequence $K^{(3)} \subseteq K^{(2)} \subseteq K^{(1)}  \subseteq K^{(0)}$ for the magnetic point groups $mmm$, $m'm'm$, $mmm'$, and $m'm'm'$. The Bulk classification for the other magnetic point groups is the same as that for the magnetic point group related by the permutation of $x,y,z$. } \label{tab:sequence}
		\begin{tabular}{cccc}
            \hline \hline
		  $M$  & IR & Pairings &  $K^{(3)} \subseteq K^{(2)} \subseteq K^{(1)}  \subseteq K^{(0)}$
			\\
			\hline    
            $mmm$  & $A_g$ & $A_g + i A_g $ & $0 \subseteq 0\subseteq 0 \subseteq 0$
            \\
            $mmm$  & $B_{1g}$ & $B_{1g} + i B_{1g}$& $0 \subseteq 0\subseteq \mathbb{Z}^2 \subseteq \mathbb{Z}^2$
            \\
            $mmm$  & $B_{2g}$ & $B_{2g} + i B_{2g} $ & $0 \subseteq 0\subseteq \mathbb{Z}^2 \subseteq \mathbb{Z}^2$
            \\
            $mmm$ & $B_{3g}$ & $B_{3g} + i B_{3g}$ & $0 \subseteq 0\subseteq \mathbb{Z}^2 \subseteq \mathbb{Z}^2$
            \\
            $mmm$  & $A_u$ & $A_u + i A_u$ & $2\mathbb{Z} \subseteq \mathbb{Z} \subseteq \mathbb{Z}^4 \subseteq \mathbb{Z}^4$
            \\
            $mmm$  & $B_{1u}$ & $B_{1u} + i B_{1u}$ & $2\mathbb{Z} \subseteq \mathbb{Z} \subseteq \mathbb{Z}^2 \subseteq \mathbb{Z}^2$
            \\ 
            $mmm$  & $B_{2u}$ & $B_{2u} + i B_{2u}$  & $2\mathbb{Z} \subseteq \mathbb{Z} \subseteq \mathbb{Z}^2 \subseteq \mathbb{Z}^2$
            \\
            $mmm$  & $B_{3u}$ & $B_{3u} + i B_{3u}$ & $2\mathbb{Z} \subseteq \mathbb{Z} \subseteq \mathbb{Z}^2 \subseteq \mathbb{Z}^2$
            \\
            $m'm'm$ & $A_g$ & $A_g + i B_{1g}$ &  $0 \subseteq 0\subseteq 0 \subseteq 0$
            \\
            $m'm'm$  & $B_g$ & $B_{2g} + i B_{3g}$ & $0 \subseteq 0\subseteq \mathbb{Z} \subseteq \mathbb{Z}$
            \\
             $m'm'm$  & $A_u$ & $A_u + i B_{1u}$ & $\mathbb{Z}  \subseteq \mathbb{Z}^3  \subseteq \mathbb{Z}^4 \subseteq \mathbb{Z}^4$
            \\
            $m'm'm$  & $B_u$ & $B_{2u} + i B_{3u}$  & $2\mathbb{Z}  \subseteq \mathbb{Z}  \subseteq \mathbb{Z} \subseteq \mathbb{Z}$
            \\
            $mmm'$  & $A_1$ & $A_g + i B_{1u}$  & $0 \subseteq 0\subseteq 0 \subseteq 0$
            \\
            $mmm'$  & $A_2$  & $B_{1g} + i A_u$ & $0 \subseteq \mathbb{Z}^2  \subseteq \mathbb{Z}^4 \subseteq \mathbb{Z}^4$
            \\
            $mmm'$&  $B_1$& $B_{2g} + i B_{3u}$ & $0 \subseteq 0\subseteq \mathbb{Z} \subseteq \mathbb{Z}$
            \\ 
            $mmm'$  & $B_2$ & $B_{3g} + i B_{2u}$ & $0 \subseteq 0\subseteq \mathbb{Z} \subseteq \mathbb{Z}$
            \\
            $m'm'm'$  & $A$  & $A_g + i A_u$  & $0 \subseteq 0\subseteq 0 \subseteq 0$
            \\
            $m'm'm'$  & $B_1$  & $B_{1g} + i B_{1u}$ & $0 \subseteq 0\subseteq 0 \subseteq 0$
            \\
            $m'm'm'$  & $B_2$  & $B_{2g} + i B_{2u}$  & $0 \subseteq 0\subseteq 0 \subseteq 0$
            \\
            $m'm'm'$ & $B_3$  & $B_{3g} + i B_{3u}$ & $0 \subseteq 0\subseteq 0 \subseteq 0$
		  \\
            \hline \hline
 \end{tabular}	
\end{table}

As an example, we consider the $A_u$ IR of $mmm$ symmetry, where the generators of $mmm$ symmetry are $\hat{2}_z$, $\hat{2}_z$, and $\hat{m}_z$, and the commutation relations are given by
\begin{align}
    &\{\hat{2}_z,\hat{2}_x\}=\{\hat{2}_x,\hat{m}_z\}=[\hat{2}_z,\hat{m}_z]=0, \nonumber \\
    &[\hat{C},\hat{2}_z]=[\hat{C},\hat{2}_x]=\{\hat{C},\hat{m}_z\}=0, \label{eq:comm_auv1}
\end{align}
where $\hat{2}_z^2=\hat{2}_x^2=\hat{m}_z^2=-1$. First, we calculate $K^{(0)}$ of the 3D Dirac Hamiltonian. From Eq.~(\ref{eq:comm_K0}), the on-site unitary operators are given by
\begin{align}
    \tilde{2}_{z,o} = \Gamma_1 \Gamma_2 \hat{2}_z, \ \ \tilde{2}_{x,o} = \Gamma_2 \Gamma_3 \hat{2}_x, \ \ \tilde{m}_{z,c}=\Gamma_3 \hat{m}_z, \label{eq:au_opv1}
\end{align}
and the commutation relations~(\ref{eq:comm_auv1}) change to
\begin{align}
     &[\tilde{2}_{z,o},\tilde{2}_{x,o}]=[\tilde{2}_{x,o},\tilde{m}_{z,c}]=[\tilde{2}_{z,o},\tilde{m}_{z,c}]=0, \nonumber \\
    &[\hat{C},\tilde{2}_{z,o}]=[\hat{C},\tilde{2}_{x,o}]=\{\hat{C},\tilde{m}_{z,c}\}=0, \label{eq:comm_auv2}
\end{align}
where $\tilde{2}_{z,o}^2=\tilde{2}_{x,o}^2=\tilde{m}_{z,c}^2=1$. Since $\tilde{2}_{z,o}$ commutes with $\tilde{2}_{x,o}$, the 3D Dirac Hamiltonian is block-diagonalized as $H_{\rm D} \to \text{diag}(h_{++},h_{+-},h_{-+},h_{--})$, where $h_{\mu \nu}$ is a matrix in the eigenspace of $\tilde{2}_{z,o} = \mu$ and $\tilde{2}_{x,o} = \nu$. Under the commutation relations~(\ref{eq:comm_auv2}), each block is not related to each other, and $\hat{C}$ and $\tilde{m}_{z,c}$ are preserved for every block, resulting in each block belonging to class DIII, where the combination of $\hat{C}$ and $\tilde{m}_{z,c}$ gives a time-reversal operator squared to $-1$. The 3D topological invariant of class DIII is $\mathbb{Z}$. Therefore, $K^{(0)} = \mathbb{Z}^4$. Let $P_{\mu \nu}$ be the projection operator onto each block. The topological invariant $\mathcal{N}_{\mu \nu}$ for each block reads~\cite{Shiozaki04A104}
\begin{align}
    \mathcal{N}_{\mu \nu} = \frac{1}{4} \text{Tr}[P_{\mu \nu} \Gamma_0 \Gamma_1 \Gamma_2 \Gamma_3 \tilde{m}_{z,c} ].
\end{align}

Next, we calculate $K^{(3)}$ that is equivalent to the classification of atomic limit phases. We consider the 3D Dirac Hamiltonian $H_{\rm D}^{O_3}$ with the 3D real representation $O_3(2_z) = \text{diag}(-1,-1,1)$, $O_3(2_x) = \text{diag}(1,-1,-1)$, and $O_3(m_z) = \text{diag} (1,1,-1)$. Using the relation (\ref{eq:comm_Kn}), the on-site unitary operations are constructed as
\begin{align}
     &\tilde{2}_{z,o}^{O_3} = i M_1 M_2 \Gamma_1 \Gamma_2 \hat{2}_z, \ \ \tilde{2}_{x,o}^{O_3} = iM_2 M_3\Gamma_2 \Gamma_3 \hat{2}_x, \nonumber\\ 
     &\tilde{m}_{z,o}^{O_3}=M_3 \Gamma_3 \hat{m}_z, \label{eq:au_opv2}
\end{align}
and the commutation relations read
\begin{align}
    &\{\tilde{2}_z^{O_3},\tilde{2}_x^{O_3}\}=\{\tilde{2}_x^{O_3},\tilde{m}_z^{O_3}\}=[\tilde{2}_z^{O_3},\tilde{m}_z^{O_3}]=0, \nonumber \\
    &\{\hat{C},\tilde{2}_z^{O_3}\}=\{\hat{C},\tilde{2}_x^{O_3}\}=[\hat{C},\tilde{m}_z^{O_3}]=0. \label{eq:comm_auv3} 
\end{align}
Since $[\tilde{2}_z^{O_3},\tilde{m}_z^{O_3}]=0$, we can block-diagonalize the 3D Dirac Hamiltonian as  $H_{\rm D} \to \text{diag}(h_{++},h_{+-},h_{-+},h_{--})$, with $h_{\mu \nu}$ belonging to the eigenspace of $\tilde{2}_{z,o}^{O_3} = \mu$ and $\tilde{m}_{z,o}^{O_3} = \nu$. In addition, $\tilde{2}_x^{O_3}$ relates $h_{\mu \nu}$ to $h_{-\mu -\nu}$, and $h_{+\nu}$ and $h_{-\nu}$ are interchanged under PH symmetry. Thus, we have a single matrix, say, $h_{++}$, which has no symmetry and belongs to class A. The topological invariant of the Dirac Hamiltonian with $n$ defect coordinates is classified by the $(3-n)$th homotopy group~\cite{Teo2010}. That is, a bound state localized at the point group center is classified by the $0$D topological invariant. The $0$D topological invariant of class A is $K_{O_3} = \mathbb{Z}$.

The image of the map $K_{O_3} \to K^{(0)}$ is obtained by omitting $M_1, M_2, M_3$ and calculating the topological invariant $\mathcal{N}_{\mu \nu}$ for generators of $K_{O_3}$. To do this, we first construct $\Gamma_i$ and $M_i$ explicitly, which read
\begin{align}
    &\Gamma_0 = \sigma_3 \tau_3 \rho_3 \omega_3, \nonumber \\
    &(\Gamma_1,\Gamma_2,\Gamma_3) = (\tau_1,\tau_3 \rho_1, \tau_3 \rho_3 \omega_1), \\
    &(M_1,M_2,M_3) = (\tau_2, \tau_3 \rho_2, \tau_3 \rho_3 \omega_2), \nonumber
\end{align}
with the on-site unitary operators $\tilde{2}_{z,o}^{O_3} = \mu_3$, $\tilde{2}_{x,o}^{O_3} = \mu_1$, and $\tilde{m}_{z,o}^{O_3} = \mu_3 \sigma_3$ and the PH operator $\hat{C}= \mu_2 \sigma_2 K$. Here, $\mu_i$, $\sigma_i$, $\tau_i$, $\rho_i$, and $\omega_i$ are independent Pauli matrices. The map $K_{O_3} \to K^{(0)}$ is calculated by omitting the mass terms $M_1$, $M_2$, and $M_3$ and evaluating $\mathcal{N}_{\mu \nu}$ in the mapped Dirac Hamiltonian. Thus, we transform $\tilde{2}_{z,o}^{O_3}$, $\tilde{2}_{x,o}^{O_3}$, and $\tilde{m}_{z,o}^{O_3}$ to $\tilde{2}_{z,o}$, $\tilde{2}_{x,o}$, and $\tilde{m}_{z,c}$ using Eqs.~(\ref{eq:au_opv1}) and (\ref{eq:au_opv2}), which leads to $\tilde{2}_{z,o}=-\mu_3 \tau_1 \rho_2$, $\tilde{2}_{x,o} = -\mu_1 \rho_1 \omega_2 $, and $\tilde{m}_{z,c} = \mu_3 \sigma_3 \tau_3 \rho_3 \omega_2$. Calculating the topological invariant $\mathcal{N}_{\mu \nu}$ yields
\begin{align}
    (\mathcal{N}_{++},\mathcal{N}_{+-},\mathcal{N}_{-+},\mathcal{N}_{--}) = (2,2,-2,-2).
\end{align}
Therefore, we obtain $K^{(3)} = 2 \mathbb{Z}$.

Finally, we consider $K^{(2)}$ that corresponds to the classification of third-order boundary states. We need to take into account all 2D representations of two mass terms. 
We here consider the subgroups of $O_3$, i.e., we omit one of the three mass terms as $x M_1 + y M_2$, $y M_2 + z M_3$, and $x M_1 + z M_3$, which we label $H_{\rm D}^{O_2}$, $H_{\rm D}^{O_2'}$, and $H_{\rm D}^{O_2''}$, respectively. Although other representations are possible, we find that the classifying groups generated from these Dirac Hamiltonian lead to the classification that is consistent with the boundary classification. We here focus on $H_{\rm D}^{O_2}$, since the other Dirac Hamiltonians lead to the same result. Since the 2D real representation is represented as $O_3(2_z) = \text{diag}(-1,-1)$, $O_3(2_x) = \text{diag}(1,-1)$, and $O_3(m_z) = \text{diag} (1,1)$, the on-site unitary operators read
\begin{align}
     &\tilde{2}_{z,o}^{O_2} = i M_1 M_2 \Gamma_1 \Gamma_2 \hat{2}_z, \ \ \tilde{2}_{x,c}^{O_2} = iM_2 \Gamma_2 \Gamma_3 \hat{2}_x, \nonumber\\ 
     &\tilde{m}_{z,c}^{O_2}= \Gamma_3 \hat{m}_z. \label{eq:au_opv3}
\end{align}
For convenience, we transform Eq.~(\ref{eq:au_opv3}) to 
\begin{align}
     &\tilde{2}_{z,o}^{O_2} = i M_1 M_2 \Gamma_1 \Gamma_2 \hat{2}_z, \ \ \tilde{m}_{y,o}^{O_2} = M_2 \Gamma_2 \hat{2}_x \hat{m}_z, \nonumber\\ 
     &\tilde{m}_{z,c}^{O_2}= \Gamma_3 \hat{m}_z, \label{eq:au_opv4}
\end{align}
where $\tilde{m}_{y,o}^{O_2} = i \tilde{2}_{x,c}^{O_2} \tilde{m}_{z,c}^{O_2}$. The commutation relations in terms of the operators in Eq.~(\ref{eq:au_opv4}) are given by
\begin{align}
    &\{\tilde{2}_{z,o}^{O_2},\tilde{m}_{y,o}^{O_2} \} = [\tilde{2}_{z,o}^{O_2},\tilde{m}_{z,c}^{O_2}]=\{\tilde{m}_{y,o}^{O_2}, \tilde{m}_{z,c}^{O_2}\}=0, \nonumber \\
     &\{\hat{C},\tilde{2}_{z,o}^{O_2}\}=[\hat{C},\tilde{m}_{y,o}^{O_2}]=\{\hat{C},\tilde{m}_{z,c}^{O_2}\}=0. \label{eq:comm_auv3} 
\end{align}
Diagonalizing the Dirac Hamiltonian according to the eigenvalue of $\tilde{2}_{z,o}^{O_2} = \pm$ leads to two blocks $h_{\pm}$, which are related by either $\hat{C}$ or $\tilde{m}_{y,o}^{O_2}$. Thus, we have a single matrix $h_+$, preserving effective PH and TR symmetries, $(\hat{C}\tilde{m}_{y,o}^{O_2})^2=1$ and $(\tilde{m}_{z,c}^{O_2}\hat{C}\tilde{m}_{y,o}^{O_2})^2=1$. Thus, $h_+$ belongs to class BDI. The classification of the Dirac Hamiltonian with a line defect ($n=2$) is the same as the 1D topological classification. We find $K_{O_2} = \mathbb{Z}$ because the 1D topological classification of class BDI is $\mathbb{Z}$. 

We explicitly construct $H_{\rm D}^{O_2}$ to find the image of $K_{O_2} \to K^{(0)}$. The matrices $\Gamma_{i}$ and $M_i$ are given by
\begin{align}
    &\Gamma_0 = \sigma_2, \nonumber \\
    &(\Gamma_1,\Gamma_2,\Gamma_3) = (\sigma_1 \tau_1, \sigma_1 \tau_3, \sigma_1 \tau_2 \rho_2), \\
    &(M_1,M_2) = (\sigma_1 \tau_2 \rho_1, \sigma_1 \tau_2 \rho_3), \nonumber
\end{align}
with the PH operator $\mu_1 K$ and the on-site unitary operators $\tilde{2}_{z,o}^{O_2} = \mu_3$, $\tilde{m}_{y,o}^{O_2} = \mu_2$, and $\tilde{m}_{z,c}^{O_2} = \mu_3 \sigma_3$. To calculate $\mathcal{N}_{\mu \nu}$, we omit $M_1$, $M_2$ and transform $\tilde{2}_{z,o}^{O_2}$, $\tilde{m}_{y,o}^{O_2}$, and $\tilde{m}_{z,c}^{O_2}$ to $\tilde{2}_{z,o}$, $\tilde{2}_{x,o}$, and $\tilde{m}_{z,c}$ using Eqs.~(\ref{eq:au_opv1}) and (\ref{eq:au_opv4}), which yields  $\tilde{2}_{z,o} = \mu_3\rho_2$, $\tilde{2}_{x,o} = \mu_1 \sigma_2 \tau_2 \rho_3$, and $\tilde{m}_{z,c} = \mu_3 \sigma_3 $. Calculating $\mathcal{N}_{\mu \nu}$ reads
\begin{align}
    (\mathcal{N}_{++},\mathcal{N}_{+-},\mathcal{N}_{-+},\mathcal{N}_{--}) = (1,1,-1,-1).
\end{align}
Therefore, we obtain $K^{(2)} = \mathbb{Z}$, which satisfies $K^{(3)} \subseteq K^{(2)}$. The bulk classification of the subgroup structures is summarized in Table~\ref{tab:sequence}.

\section{Boundary classification}
\label{app:boundary}

The classifying group $\mathcal{K}_a^{(n)}$ classifies $n$th-order boundary states when the crystal shape is compatible with the magnetic point group symmetry. In this Appendix, we explain the boundary classification approach discussed in Ref.~\cite{Luka011012,Luka202000090}, where they used the $M$-symmetric cellular decomposition of a crystal. Note that the classification using the cellular decomposition is also discussed in Refs. ~\cite{Kruthoff2017,ZSong2019,Okuma2019,Shiozaki2022,Shiozaki2023Generalized,Shiozaki01827}. Let $X$ be the interior of a $d$ dimensional crystal, which can be decomposed as
\begin{align}
    X= \Omega_0 \cup \Omega_1 \cup \cdots \cup \Omega_d,
\end{align}
where $\Omega_k$ is a set of disjoint $k$ cells $c_k$, which are a $k$ dimensional subspace of $X$ that is homotopic to a $k$ dimensional sphere. The $M$-symmetric cellular decomposition satisfies that an element of the magnetic point group $M$ acts on each cell as on-site symmetry or it moves the cell to different cells. For each $k$ cell, we consider a $k$ dimensional topological phase classified by the classifying group with on-site symmetries that leave $c_k$ invariant, which has a $(k-1)$ dimensional boundary state on its boundary $\partial c_k$. By placing a $k$-dimensional topological phase on each $k$ cell in a $M$-symmetric manner and assuming that the boundary states that arise in the interior of $X$ gap out, we can generate all possible $(k-1)$-dimensional boundary state on the boundary between $\Omega_k$ and $X$, $\partial \Omega_k \cap \partial X$. Note that the boundary states include both intrinsic and extrinsic states. By setting $k=d+1-n$, we denote $\mathcal{K}^{(n)}$ as the classifying group of all $n$th-order topological boundary states on $\partial \Omega_{d+1-n} \cap \partial X$. 

To find the classifying group $\mathcal{K}_a^{(n)}$, we need to separate intrinsic and extrinsic boundary states. We refer $\mathcal{D}^{(n)} \subset \mathcal{K}^{(n)}$ as the classifying group for the extrinsic boundary states, which reside only within $\partial X$. The classifying group $\mathcal{D}^{(n)}$ can be obtained by the decoration method as follows. The crystal boundary $\partial X$ can be decomposed as
\begin{align}
    \partial X= \Omega_0^{\partial} \cup \Omega_1^{\partial} \cup \cdots \cup \Omega_d^{\partial},
\end{align}
where $\Omega_k^{\partial} = \Omega_{k+1} \cap \partial X$, referring to all possible $k$-dimensional boundaries of the crystal $X$.
Similarly to the construction of $k$ dimensional topological phases in the interior of $X$, $\mathcal{D}^{(n)}$ can be obtained by putting topological phases on $k$ cells $\Omega^{\partial}_{k}$ with $k \ge d+1-n$, where all states of dimension $>d-n$ can be gapped out.  Using $\mathcal{D}^{(n)}$, the classifying group of intrinsic boundary states is given by
\begin{align}
    \mathcal{K}_a^{(n)} = \mathcal{K}^{(n)}/\mathcal{D}^{(n)}.
\end{align}

For 3D TRSB superconductors, $ \mathcal{K}_a^{(1)} =0$ since there is no first-order topological phase. For the magnetic point groups $mmm$, $m'm'm$, $mmm'$, and $m'm'm'$ and the related groups, lower-dimensional topological invariants are given by the Chern numbers, mirror Chern numbers, 1D $\mathbb{Z}_2$ topological invariants, and 1D magnetic winding numbers as shown in Table~\ref{tab:topology}, which are related to a surface state that constitutes the classifying group $\mathcal{K}^{(n)}$. 
Lower dimensional TSCs are pasted on $\Omega_k$ ($\Omega^{\partial}_k$) in a $M$-symmetric way to obtain $\mathcal{K}^{(n)}$ ($\mathcal{D}^{(n)}$). At the intersection between two $k$-cells, we check whether there exists a symmetry-preserving mass term that gaps out the boundary state of the pasted TSCs. If the boundary states are gapped, the configuration constitutes an element of $\mathcal{K}^{(n)}$. In contrast, for the surface decoration, the boundary states that remain stable constitutes an element of $\mathcal{D}^{(n)}$.

\begin{figure}[t]
\centering
    \includegraphics[scale=0.061]{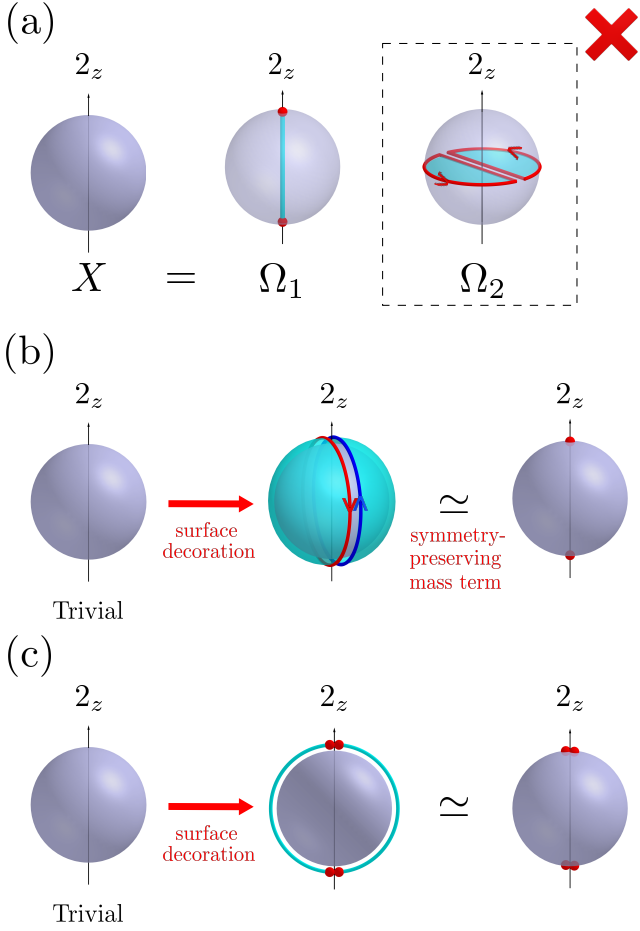}
    \caption{Schematic illustration of bulk cellular decomposition (a) and surface decorations [(b),(c)] for the case of the magnetic point group $M=2$ and $B$ pairing state. The gray sphere represents a 3D system that is compatible with the magnetic point group symmetry. In (b) and (c), we start from trivial states that have no boundary mode, and then construct an extrinsic topological surface state by pasting 2D (1D) TSCs on each cell. The pasted  topological states are illustrated by the blue places (lines). The chiral edge modes of 2D TSCs are depicted by the red and blue arrows, whose direction indicates the propagating direction of chiral edge modes. The edge states of the 1D TSCs are denoted by the red points. If there is a symmetry-preserving mass term, the configuration of boundary states change.}
    \label{fig: Decoration-rotation}
\end{figure}

To understand the essence of the boundary classification, we consider a system with the magnetic point group symmetry $M=2=\{e,2_z\}$ and $B$ pairing state, i.e., $\eta_{2_z}=-1$. From the bulk classification, the classifying group of anomalous surface states are given by~\cite{Luka011012} 
\begin{align}
    \mathcal{K}_a^{(1)}=0, \ \ \mathcal{K}_a^{(2)}=0, \ \ \mathcal{K}_a^{(3)}=0. \label{eq:Ka-2B}
\end{align}
In the following, we revisit this result from the boundary classification.
First, we consider the bulk cellular decomposition of bulk crystal $X$, where $X$ is chosen to be a sphere for simplicity. The bulk crystal $X$ can be decomposed into $k$-dimensional cells $\Omega_k$ ($k=0,1,2,3$), and then, $k$-dimensional TSCs are pasted on each cell. Hereafter, we omit $\Omega_0$ since it is irrelevant to boundary states. Boundary states arise from Majorana zero energy states at the edge of 1D TSCs ($\partial c_1$) and Majorana chiral edge states at the boundary of 2D TSCs ($\partial c_2$). To obtain a fully-gapped state, boundary states that appear in the interior of $X$ must be gapped out.

Second-order boundary states are constructed by putting 2D TSCs on two $2$-cells that are interchanged by $2_z$ as shown in Figure~\ref{fig: Decoration-rotation} (a). Since there is no on-site symmetry, it belongs to class D, and topological state have a chiral Majorana edge mode on its boundary. To find out whether the interior of $X$ is gapped out, we consider a 1D Dirac Hamiltonian describing two chiral edge modes in the intersection of the two cells as 
\begin{align}
H_{2_z}(k_1) = k_1 s_x, \label{eq:hd_2z_int}
\end{align}
which satisfy
\begin{align}
 &\hat{2}_z H_{2_z}(k_1) \hat{2}^{-1}_z = H_{2_z}(-k_1), \ \ \hat{2}_z = -is_z, \nonumber \\
 &\hat{C} H_{2_z}(k_1) \hat{C}^{-1} = -H_{2_z}(-k_1), \ \ \hat{C}=K, \nonumber  
\end{align}
where the basis of Pauli matrices $s_i$ denotes two chiral edge states, and $k_1=i \partial_{x_1} \ (x_1 \perp z)$ is a momentum in the direction of the chiral edge modes. We find that Eq. (\ref{eq:hd_2z_int}) has no mass term $M_0$ that satisfies $\{M_0, H_{2_z}(k_1)\}=\{M_0, \hat{C}\}=[M_0,\hat{2}_z]=0$. Thus, we obtain $\mathcal{K}^{(2)}=0$. 

Similarly, we consider third-order boundary states, which are constructed by putting 1D TSCs in a $1$-cell on the rotation axis as shown in Figure~\ref{fig: Decoration-rotation} (a). Since the $1$ cell is symmetric under the $2_z$ rotation, the 1D Hamiltonian is block-diagonalized into two blocks in terms of the eigenvalues of $\hat{2}_z=\pm i$. Moreover, the commutation relation between the PH and $2_z$ operators satisfy $\eta_{2_z}=-1$, so that each block belongs to class D. Thus, Majorana zero energy states at the edges are classified by $\mathcal{K}^{(3)}=\mathbb{Z}_2 \oplus \mathbb{Z}_2$. 

We now consider $\mathcal{D}^{(3)}$ to find the intrinsic topological surface state. $\mathcal{D}^{(3)}$ can be constructed from the topological states placed on $\Omega^{\partial}_1$ and $\Omega^{\partial}_2$. For $\Omega^{\partial}_2$, the possible surface decoration is depicted in Figure~\ref{fig: Decoration-rotation} (b), where there are two 2-cells that are related by $2_z$.  Putting a 2D TSC with a chiral edge mode on one of the 2-cells, the other must have a chiral edge mode in the opposite direction due to $2_z$ symmetry. To check the stability of chiral edge modes, we consider a 1D Dirac Hamiltonian describing the two chiral edge modes on the boundary as 
\begin{align}
H^{2_z}_{\text{bdry}}(k_1) = k_1 s_x, \label{eq:h1dD_2B}
\end{align}
with
\begin{align}
 &\hat{2}_z H^{2_z}_{\text{bdry}}(k_1) \hat{2}^{-1}_z = H^{2_z}_{\text{bdry}}(-k_1), \ \ \hat{2}_z = -is_z, \nonumber \\
 &\hat{C} H^{2_z}_{\text{bdry}}(k_1) \hat{C}^{-1} = -H^{2_z}_{\text{bdry}}(-k_1), \ \ \hat{C}=K, \nonumber   
\end{align}
in a similar way to Eq.~(\ref{eq:hd_2z_int}). Equation (\ref{eq:h1dD_2B}) then allows a mass term as
\begin{align}
M_1(x_1) = f(x_1) s_y, \quad f(-x_1)=-f(x_1),
\end{align}
which gap out the chiral edge modes except for $x_1=0$, indicating that a pair of Majorana corner states appears as the Jackiw-Rebbi zero-energy solutions~\cite{Jackiw1976} at $x_1=0$, i.e., on the rotation axis. This configuration corresponds to the case of $(1,0)$ or $(0,1) \in \mathbb{Z}_2 \oplus\mathbb{Z}_2$. On the other hand, the surface decoration on $\Omega^{\partial}_1$ is illustrated in Figure~\ref{fig: Decoration-rotation} (c), where we have two 1-cells that are related by the $2_z$ operation. Pasting the 1D TSCs on each 1-cell lead to two pairs of the Majorana corner states on the rotation axis. Note that a 1D Hamiltonian in the $1$-cells is not invariant under the $2_z$ operation. Since there is no symmetry-preserving mass term due to the protection from $2_z$ symmetry, these corner states remain stable. This configuration corresponds to the case of $(1,1) \in \mathbb{Z}_2 \oplus\mathbb{Z}_2$. Therefore, we conclude $\mathcal{D}^{(3)}=\mathbb{Z}_2 \oplus \mathbb{Z}_2$ and $\mathcal{K}^{(3)}_a=\mathcal{K}^{(3)}/\mathcal{D}^{(3)}=0$. The result aligns with the classifying group (\ref{eq:Ka-2B}).

In the following, we apply the boundary classification to the $B_{2u} + iB_{3u}$ and $A_u + iA_u$ pairing states as concrete examples.

\subsection{$B_{2u} + iB_{3u}$ pairing}

\begin{figure}[t]
\centering
    \includegraphics[scale=0.061]{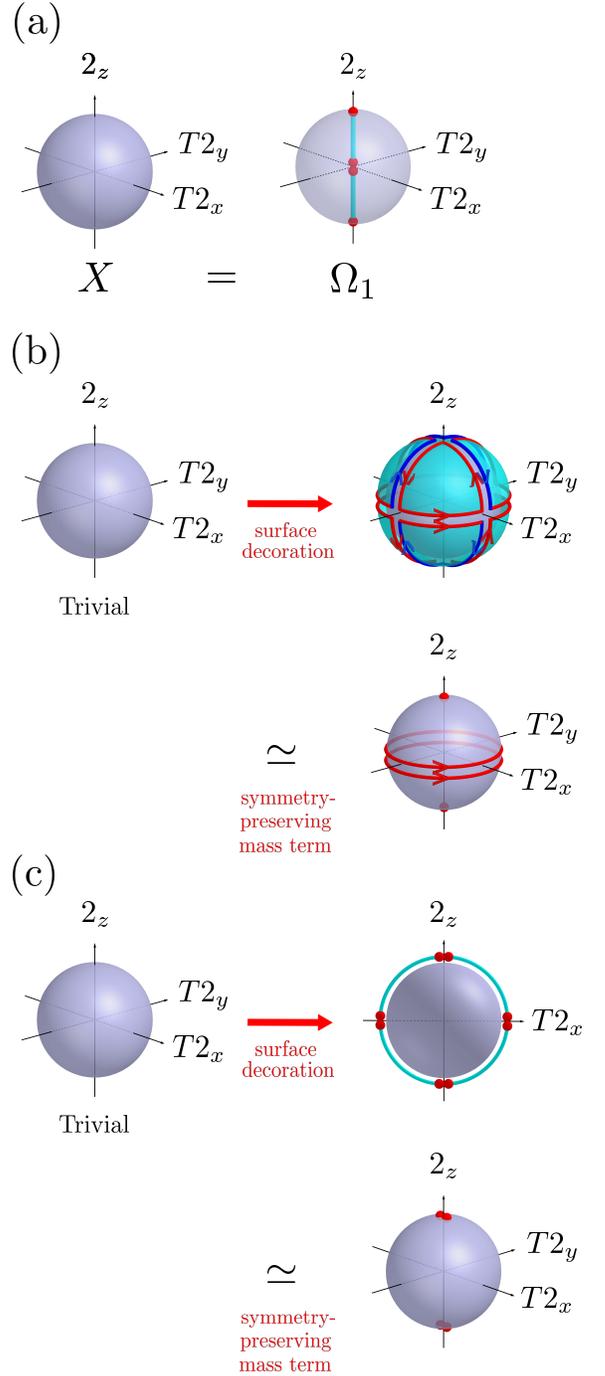}
    \caption{Schematic illustration of the surface decorations for the case with the magnetic point group $M=m'm'm$ and $B_u$ pairing state. The symbols are the same as those in Figure \ref{fig: Decoration-rotation}. 
    }
    \label{fig: surface-decoration-m'm'm}
\end{figure}

We consider the $B_{2u} + iB_{3u}$ pairing state, which preserves $m'm'm$ symmetry and satisfies $-\eta_{2_z}=\eta_{m_z}=-\eta_{I}=1$. From Table~\ref{tab:pairings}, the classifying group of anomalous surface states is obtained as
\begin{align}
    \mathcal{K}_a^{(1)}=0, \ \ \mathcal{K}_a^{(2)}=0, \ \ \mathcal{K}_a^{(3)}=\mathbb{Z}_2.
\end{align}
To understand this result from the boundary classification, we construct $\mathcal{K}^{(n)}$ as shown in Figure~\ref{fig: surface-decoration-m'm'm} (a). For $\Omega_2$, we readily find $\mathcal{K}^{(2)}=0$ since there is no 2D TSC; on the $(xy)$ plane, the BdG Hamiltonian block-diagonalized in eigenspaces of $m_z = \pm i$ belongs to class A since $\eta_{m_z}=1$. In class A, there is no Majorana chiral edge mode. On the $(yz)$ and $(zx)$ planes, the BdG Hamiltonian belongs to class BDI due to $(T2_i)^2=1$ $(i=x,y)$, which does not have any stable 2D topological phase. For $\Omega_1$, there are two cases: pasting a 1D TSC on the $2_z$ symmetric axis and one on the $T2_i \ (i=x,y)$ symmetric axis. In the former case, 1D TSCs are classified by $\mathbb{Z}_2 \oplus \mathbb{Z}_2$ since the BdG Hamiltonian satisfies $2_z$ symmetry and $\eta_{2_z}=-1$. We can put two 1D TSCs on the $2_z$ symmetric axis, which are related by inversion symmetry and intersect at the inversion center. The edge states of each 1D TSC in the interior of $X$ can be gapped out by the symmetry-preserving mass term $M_0=s_z$ that satisfies the symmetries $2_z=-i$, $m_z=is_z$, $T2_x = K$, and $C=s_x K$, where the basis of the Pauli matrices denotes the edge modes at the inversion center. Thus, we can past this 1D TSC without any obstruction. In the letter case, a 1D TSC belongs to class AI since the BdG Hamiltonian is invariant under $T2_i \ (i=x,y)$ and $m_z$ symmetries. Hence, it splits into two matrices in terms of eigenspaces of $m_z$, and each eigenspace only preserves $(T2_i)^2=1$. Thus, there is no 1D TSC. 
Therefore, we conclude $\mathcal{K}^{(3)}=\mathbb{Z}_2 \oplus \mathbb{Z}_2$.

\begin{figure}[t]
\centering
    \includegraphics[scale=0.07]{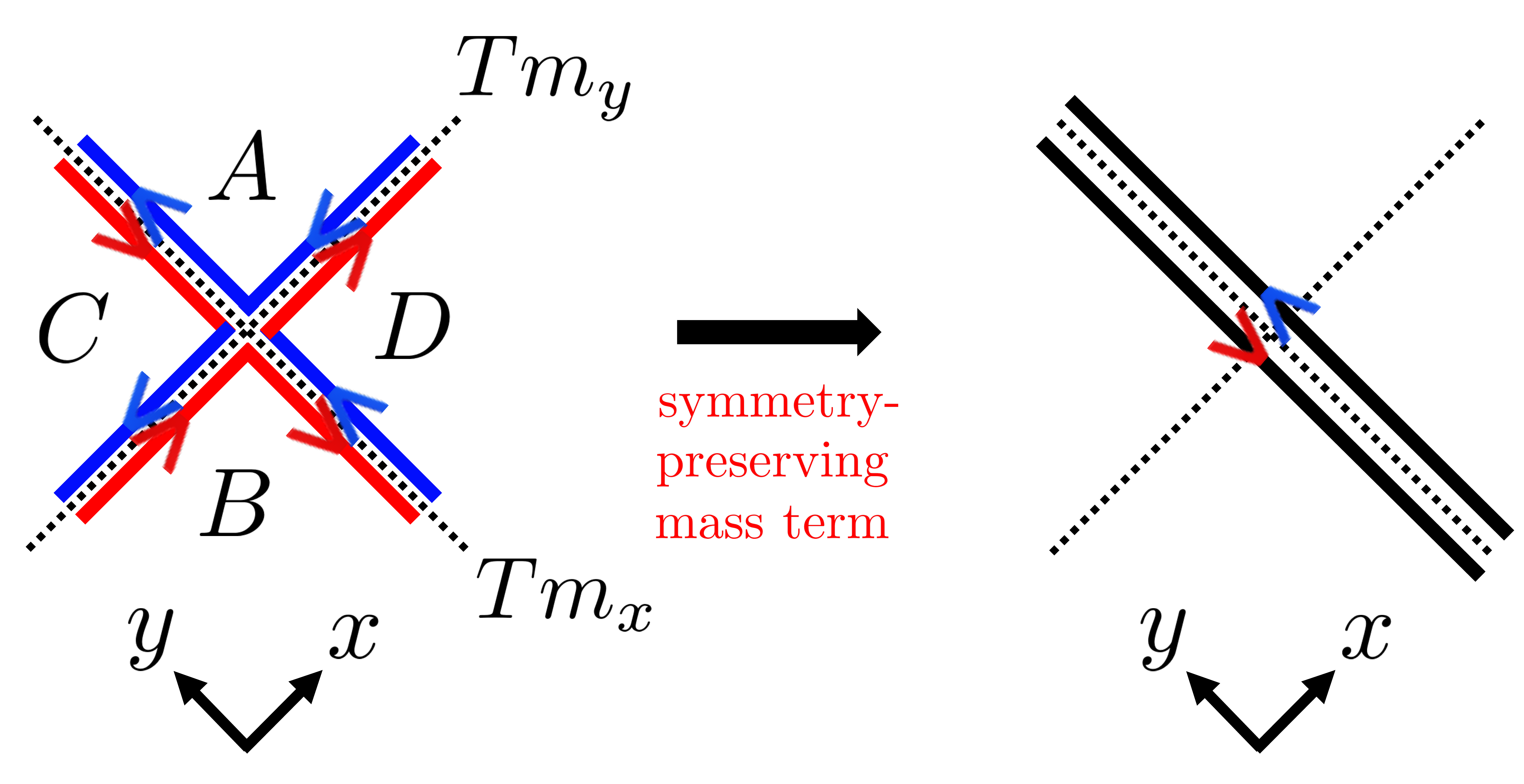}
    \caption{Schematic illustration of four chiral edge modes around the $2_z$ rotation axis, where A, B, C, D label the 2-cells. They can transform to two chiral modes by adding the symmetry-preserving mass term.}
    \label{fig: surface-decoration-m'm'm-effective}
\end{figure}

To extract intrinsic topological surface states, we construct $\mathcal{D}^{(3)}$. We first consider 2D TSCs placed on $\Omega^{\partial}_2$. The possible configuration is shown in Figure \ref{fig: surface-decoration-m'm'm} (b), where we have eight 2-cells on the sphere, and the 2D TSCs are put on each 2-cells in such a way that the direction of the chiral edge modes is compatible with $m'm'm$ symmetry. In Figure \ref{fig: surface-decoration-m'm'm} (b), the chiral edge modes on the latitude of the sphere has a counter-propagating chiral edge mode from the neighboring 2-cells, which can be gapped out by a mass term. On the other hand, the chiral edge modes on the equator of the sphere propagate in the same direction, so that no mass term exists. To see the behavior of the chiral edge modes on the latitude, we model a 1D Dirac Hamiltonian describing the four chiral edge modes around the $2_z$ rotation axis as 
\begin{align}
    &H^{m'm'm}_{\rm bdry}(k_1) = \left[ 
\begin{array}{cccc}
  h_{\rm A}  &&& \\
  &h_{\rm B} && \\
  &&h_{\rm C}& \\
  &&& h_{\rm D}
\end{array} 
\right]_{s \otimes \tau}, \label{eq:H1dD_patch}
\end{align}
with
\begin{align*}
   & h_{\rm A} (k_1) = k_1, \ \ h_{\rm B} (k_1) = O_{2_z} k_1, \\
   & h_{\rm C} (k_1) = - O_{m_y} k_1, \ \  h_{\rm D} (k_1) = - O_{m_x} k_1,
\end{align*}
where A, B, C, and D label the patches as shown in Figure~\ref{fig: surface-decoration-m'm'm-effective} (left), and $k_1= i \partial_{x_1}$ is a momentum in the direction of the chiral edge mode in the patch A. Equation (\ref{eq:H1dD_patch}) satisfies the symmetry constraints from $\hat{C}=K$, $\hat{2}_z = -is_x$, $\widehat{Tm}_x = i\tau_x K$, and $\widehat{Tm}_y = s_x\tau_x K$. In the $x$ direction, Eq.~(\ref{eq:H1dD_patch}) reduces into
\begin{align}
    H^{m'm'm}_{\rm bdry}(k_x) = \text{diag}[-k_x, k_x, -k_x , k_x ]_{s \otimes \tau}.
\end{align}
We can find a symmetry-preserving mass term $M_1(x) = f_1(x) s_y \tau_x$ with $f_1(-x)=-f_1(x)$, which opens a gap between the chiral edge modes of the patches A and D (C and B) while the chiral edge modes in the $y$ direction remain intact. (see Figure~\ref{fig: surface-decoration-m'm'm-effective} (right)). Furthermore, the remaining chiral edge modes have a Jackiw-Rebbi zero energy state at $x=y=0$, i.e., the $2_z$ rotation axis, when a symmetry-preserving mass term is added. The mechanism is the same as the case of the B pairing state of $M=2$. 

On the other hand, $\Omega^{\partial}_1$ consists of twelve 1-cells on the sphere.
Pasting 1D TSCs on $\Omega^{\partial}_1$ is shown in Figure \ref{fig: surface-decoration-m'm'm} (c). 
In this case, we can find a symmetry-preserving mass term at two Majorana corner states on the $m_z$ mirror plane. To see this, we consider a zero energy state on the $m_z$ mirror plane, which is invariant under $\hat{C}=s_x K$, $\widehat{T2}_x=-is_zK$, $\hat{m}_z=is_z$, and $\widehat{Tm}_y = K$. We then find a symmetry-preserving mass term $M_0=s_z$, indicating that the zero energy mode is gapped out by the perturbation. That is, two Majorana corner states on the $2_z$ rotation axis only remain stable. We obtain $\mathcal{D}^{(3)}$ as in Figures~\ref{fig: surface-decoration-m'm'm} (b) and (c), resulting in that third-order boundary states characterized by $(1,1) \in \mathbb{Z}_2 \oplus \mathbb{Z}_2$ are removed by the surface decoration. Therefore, we conclude $\mathcal{K}^{(3)}_a = \mathbb{Z}_2$.  It is noteworthy that the single corner states at the rotation axis are interchangeable with the double chiral edge modes on the equator by pasting the extrinsic topological surface state.

\subsection{$A_{u} + iA_{u}$ pairing}

\begin{figure}[t]
\centering
    \includegraphics[scale=0.059]{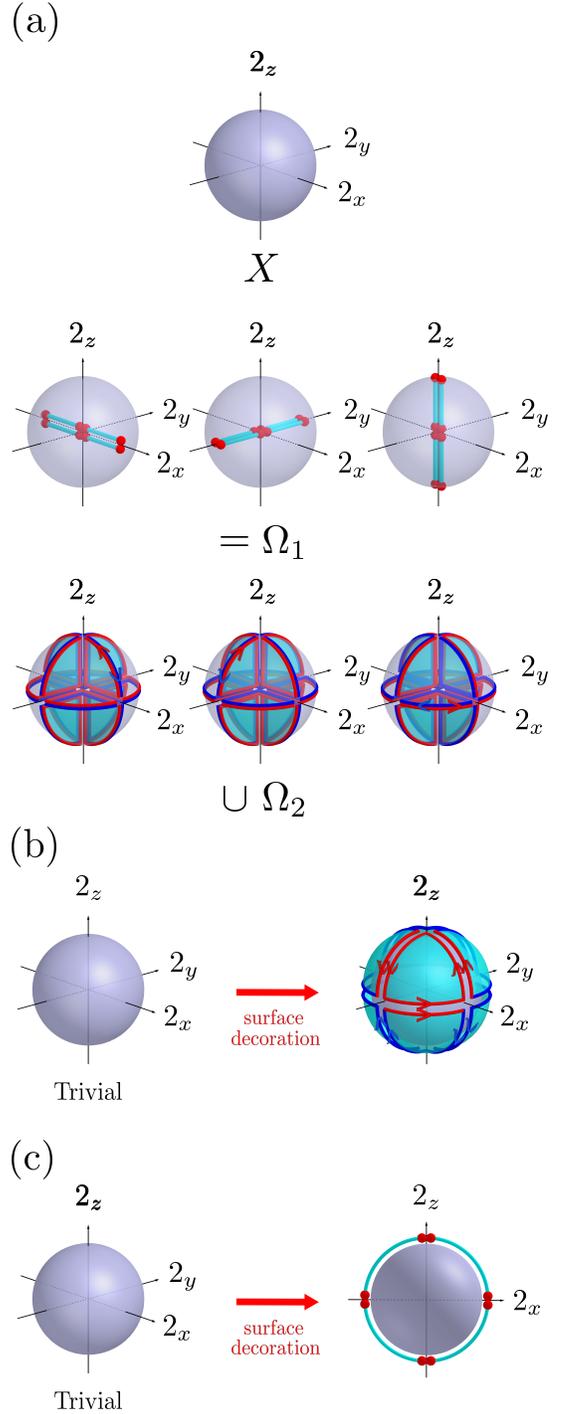}
    \caption{Schematic illustration of the surface decoration for the case of the magnetic point group $M=mmm$ and $A_u$ pairing state. The symbols are same as those in Figure~\ref{fig: Decoration-rotation}.}
    \label{fig: surface-decoration-mmm}
\end{figure}

As another example, we consider the $A_{u} + iA_{u}$ pairing state, which preserves $mmm$ symmetry and satisfies $\eta_{2_z}=\eta_{2_y}=\eta_{2_x}=-\eta_{I}=1$. 
Table~\ref{tab:pairings} reads the classifying group of anomalous surface states as 
\begin{align}
    \mathcal{K}_a^{(1)}=0, \ \ \mathcal{K}_a^{(2)}=\mathbb{Z}^3, \ \ \mathcal{K}_a^{(3)}=\mathbb{Z}_2.
\end{align}
We revisit this classification using the boundary classification as follows. 
First, we consider $\mathcal{K}^{(2)}$, which have twelve $2$-cells in the interior of $X$ that consist of four $2$-cells on each mirror plane as shown in Figure~\ref{fig: surface-decoration-mmm}. Since $\eta_{m_z}=\eta_{m_y}=\eta_{m_x}=-1$, the BdG Hamiltonian on each mirror plane belongs to class D, characterized by the mirror Chern number $\mathbb{Z}$. To check whether the interior of $X$ is gapped out, we consider chiral edge modes on the $m_z$ mirror plane, which is given by
\begin{align}
    &H_{\text{mmm}}(k_1) = \text{diag}[H^{\text{mmm}}_+,H^{\text{mmm}}_-]_{s \otimes \tau \otimes \mu}, \label{eq:H1dD_mmm}
\end{align}
with
\begin{align}
    H^{\text{mmm}}_\pm(k_1) = \pm \text{diag}[h_{\rm A},h_{\rm B},h_{\rm C},h_{\rm D}]_{s \otimes \tau},
\end{align}
where $H^{\text{mmm}}_\pm$ are the Dirac Hamiltonian describing the four chiral edge modes in the eigenspaces of $m_z=\pm i$, and $h_{\rm A-D}$ are the Hamiltonians defined in Eq.~(\ref{eq:H1dD_patch}). Equation (\ref{eq:H1dD_mmm}) satisfies $\hat{C}= K$, $\hat{2}_z=-is_y$, $\hat{m}_z = i\mu_z$, $\hat{m}_x = i s_z \tau_x \mu_x$, and $\hat{m}_y=is_x \tau_x \mu_x$. We then find a symmetry-preserving mass term $M_1 (x) = s_x \tau_y \mu_z f_1(x)$ with $f_1(-x) = - f_1(x)$, and obtain the two chiral edge modes in each mirror sector in the $y$ direction as shown in Figure~\ref{fig: surface-decoration-m'm'm-effective}. Thus, there remains a Jackiw-Rebbi zero energy state in each mirror sector in a similar way to the argument in the $B_{2u} + iB_{3u}$ pairing. Thus, we need to consider multiple mirror planes together to pair-annihilate these zero energy modes. When putting 2D TSCs on three orthogonal mirror planes in a $mmm$ symmetric way, adding a mass term $M_1 (z) \propto f_1(z)$ gaps out the chiral edge modes in the direction perpendicular to the $m_z$ mirror plane, while those in the $m_z$ mirror plane remain stable. The remaining configuration corresponds to a doubled Hamiltonian of Eq.~(\ref{eq:H1dD_mmm}), which allows a fully gapped state.   
There are eight patterns of such configurations, characterized by $(\text{Ch}_1[m_x],\text{Ch}_1[m_y],\text{Ch}_1[m_z])=(\pm 1, \pm 1, \pm 1)$, and three of them are independent of each other. 
Thus, we find $\mathcal{K}^{(2)}=\mathbb{Z}^3$. 

On the other hand, for $\Omega_1$, the $x$, $y$, and $z$ axes are invariant under $2_i$ and $I$ symmetries. The situation is the same as those in the case of $B_{2u} + i B_{3u}$ pairing, but the classification of 1D Hamiltonians change from $\mathbb{Z}_2 \oplus \mathbb{Z}_2$ to $\mathbb{Z}_2$ since $\eta_z =1$. When putting two 1D TSCs on the $2_i$ symmetric axis in an inversion symmetric way, zero energy edge modes intersect at the inversion center, which can be gapped out by a symmetry-preserving mass term $M_0=\sigma_z$ that preserves $2_z=-is_z$, $2_x = -is_x$, $m_z=is_z \sigma_z$, and $C=s_y\sigma_yK$, where the Pauli matrices $\bm{s}$ and $\bm{\sigma}$ denote the four zero energy states at the inversion center. Therefore, we conclude $\mathcal{K}^{(3)}=\mathbb{Z}_2^3$.

We now construct $\mathcal{D}^{(2)}$ and $\mathcal{D}^{(3)}$, where the cellular decomposition is the same as that of the $m'm'm$ symmetry.  
For $\Omega^{\partial}_2$, it is impossible to place the 2D TSCs in a $M$-symmetric way due to the multiple mirror-reflection symmetries. An example is shown in Figure~\ref{fig: surface-decoration-mmm} (a), which is not consistent with $mmm$ symmetry. Thus, we obtain $\mathcal{D}^{(2)}=0$. On the other hand, pasting the 1D TSCs on $\Omega^{\partial}_1$ is possible. The possible configuration is illustrated in Figure \ref{fig: surface-decoration-mmm} (b). In this case, two Majorana corner states appear at each corner and are stable since there is no symmetry-preserving mass term. For instance,  consider two zero energy modes on the $2_x$ rotation axis that preserves $\hat{2}_x=-is_x$, $\hat{m}_y=is_y$, $\hat{m}_z = is_z$, and $\hat{C}=s_zK$, under which there is no mass term. Note that even when we have two Majorana corner states, it can be classified by $\mathbb{Z}_2$ since there is a symmetry-reserving mass term $\tau_y$ in the doubled boundary Hamiltonian.
Hence, the configurations of the Majorana corner states characterized by $(1,1,0)$, $(1,0,1)$, and $(0,1,1) \in \mathbb{Z}_2^3$ are categorized as an extrinsic topological surface state. Since only two of them are independent, we obtain $\mathcal{D}^{(3)} = \mathbb{Z}_2^2$. Therefore, we conclude $\mathcal{K}^{(2)}_a = \mathbb{Z}^3$ and $\mathcal{K}^{(3)}_a = \mathbb{Z}_2$.
Interestingly, several configurations of Majorana corner states are possible through the surface decoration. For instance, the configuration with multiple Majorana corner states on all rotation axes characterized by $(1,1,1) \in \mathcal{K}^{(3)}$ can change to those with a single Majorana corner state in a rotation axis characterized by $(1,0,0)$, $(0,1,0)$, or $(0,0,1)$.

\section{Definition of topological invariants}
\label{app:invariant}
\subsection{Chern number}
In class D, a 2D topological phase is characterized by the Chern number $\text{Ch}_1 \in \mathbb{Z}$:
\begin{align}
&\text{Ch}_1 = \frac{1}{2\pi}\int_{\text{2D BZ}} d^2k \ \mathcal{F}(\bm{k}) , 
\label{eq:Chern} \\
&\mathcal{F}(\bm{k}) = [\bm{\nabla}_{\bm{k}} \times \bm{a}(\bm{k})]_{\perp}, 
\end{align}
where $\mathcal{F}(\bm{k})$ is the Berry curvature, $[\cdots]_{\perp}$ means the component perpendicular to the 2D BZ and $\bm{a}(\bm{k})$ is the Berry connection of the occupied states,
\begin{align}
    \bm{a}(\bm{k}) = -i \sum_{n \in {\rm occ}} \langle n, \bm{k}| \bm{\nabla}_{\bm{k}} | n,\bm{k}\rangle.
\end{align}
In 3D systems, the Chern number is defined in a 2D subspace of the 3D BZ. The nonzero Chern number implies the existence of superconducting point nodes, accompanying surface Majorana arc states terminating at the point nodes.  
The Chern number is zero when $TI$ symmetry or mirror-reflection symmetry whose mirror plane is perpendicular to the 2D BZ is preverved.

\subsection{Mirror Chern number}
When the pair potential satisfies $\eta_{m_i}=-1$, the mirror Chern number is defined by
 \begin{align}
 \text{Ch}_1[m_i] \equiv \frac{\text{Ch}_1^+ -\text{Ch}_1^-}{2},
\label{eq:mirror-Chern}
\end{align} 
 where $\text{Ch}_1^{\pm} \in \mathbb{Z}$ is the Chern number defined in the eigenspace of $\tilde{D}(m_i) = \pm i$ on the mirror plane. 
  The Chern number satisfies
 \begin{align}
     \text{Ch}_1 = \text{Ch}_1^+ + \text{Ch}_1^-. \label{eq:mirror-ch}
 \end{align}
 Thus, $\text{Ch}_1^+ =- \text{Ch}_1^-$ when $\text{Ch}=0$.
 When $H$ includes multiple mirror-reflection symmetries, $\text{Ch}_1^+ =- \text{Ch}_1^-$ for each mirror plane since $\{D(m_i),D(m_j)\}=0$ ($i\neq j$).

\subsection{1D magnetic winding number}
 
When $T(G-H) \neq \emptyset$, the 1D magnetic winding number is defined by, in a symmetric 1D subspace of the 3D BZ, 
\begin{align}
W[h] = \frac{i}{4\pi}\int_{-\pi}^\pi dk \tr[\Gamma(Th) H^{-1}(k)\pdv{H(k)}{k}], \label{eq:winding-1} 
\end{align}
where $Th=T2_i, Tm_i$ ($i=x,y,z$), $H(k)$ is the BdG Hamiltonian in the 1D subspace, the magnetic chiral operator is defined by $\Gamma(Th) \equiv e^{i \phi} \tilde{D}(Th) C$. The magnetic chiral operator satisfies $\Gamma^2(Th) =1$ and $\{\Gamma(Th), H(k)\}=0$. 
 $W[h] \neq 0$ requires pair potentials satisfying~\cite{xiong17}
\begin{align}
-p(h,g)\eta_g = p(h,g')\eta_{g'} = -1, \label{eq:winding-constraint}
\end{align}
where $hg=p(h,g)gh$ and $g,g' \in H$ satisfies $g k = k$ and $g' k = -k$. Equation (\ref{eq:winding-constraint}) means $W[h] = 0$ for even-parity superconductors. 

\subsection{1D $\mathbb{Z}_2$ topological invariant}

When $g \in H$ and $\eta_g = -1$, the 1D $\mathbb{Z}_2$ topological invariants in terms of $g=2_i, m_i$ ($i=x,y,z$) are defined as
\begin{align}
\nu[g]_{\pm}= \frac{1}{\pi} \int_{-\pi}^{\pi} dk \; a^{g}_{\pm}(k) \quad \text{mod $2$}, \label{rotation-invariant} 
\end{align}
where $a^{g}_{\pm}$ is the Berry connection defined in the eigenspace of $\tilde{D}(g)=\pm i$  and the integration is performed in the $g$ symmetric 1D subspace of the 3D BZ.

\subsection{Inversion symmetry indicator}
 When $I \in H$, TRSB odd-parity superconductors ($\eta_I=-1$) has an additional invariant associated with inversion symmetry $\kappa[I] \in \mathbb{Z}_8$, which is defined by~\cite{Anastasiia013064}
\begin{align} 
\kappa[I] \equiv \frac{1 }{2} \sum_{\bm{k}\in \text{TRIM}} \Big(n^{+}_{\bm{k},\text{BdG}} -n^{-}_{\bm{k},\text{BdG}}\Big) 
\end{align}
where $n^{\alpha}_{\bm{k},\text{BdG}}$ is the BdG occupied bands with inversion eigenvalue $\alpha = \pm 1$ at TRIMs~\footnote{The BdG occupied bands are defined by the difference between the number of occupied bands in the BdG Hamiltonian at $\mu = \epsilon_{\rm F}$ (the Fermi level) and that at $\mu=-\infty$.}. The summation is taken over all TRIMs in the 3D BZ.

\subsection{Inversion symmetry indicator in the mirror plane}
 When $I, m_i \in H$ and $\eta_I = \eta_{m_i} =-1$, there is $\kappa[I]$ restricted to the eigenspace of $\tilde{D}(m_i) = \pm i$, which define a $\mathbb{Z}_4$ index as 
 \begin{align}
     \kappa[I]^{m_i}_{\pm} \equiv  \frac{1 }{2} \sum_{\bm{k}\in \text{TRIM}} \Big(n^{\pm +}_{\bm{k},\text{BdG}} -n^{\pm -}_{\bm{k},\text{BdG}}\Big)  
 \end{align}
 where $n^{\pm +}_{\bm{k},\text{BdG}}$ ($ n^{\pm -}_{\bm{k},\text{BdG}}$) is  $n^{+}_{\bm{k},\text{BdG}}$ ($n^{-}_{\bm{k},\text{BdG}}$) with mirror eigenvalue $\pm i$, and the summation of momentum is taken over 2D TRIMs in the mirror plane. In the crystalline symmetry of UTe$_2$, $\kappa[I]^{m_i}_{\pm}$ appears only when the BdG Hamiltonian preserves $mmm$ symmetry. In this symmetry class, the eigenstates of $m_i=\pm i$ are related by other symmetry operators, so that $\kappa[I]^{m_i}_{+} = \kappa[I]^{m_i}_{-}$.  

\section{Relationship between topological invariants for HOTPs}
\label{app:generator}

The classifying groups have multiple topological invariants, which are related to each other. We examine the relationship between topological invariants using a generator of the classifying groups, which is described by a Dirac Hamiltonian.
To see this, we consider a 2D class D superconductor with aniunitary symmetry $T2_y$ as an example. The model is described by a 2D Dirac Hamiltonian as
\begin{align}
    H_{\rm 2d D} = M_0 \Gamma_0 + k_x \Gamma_x + k_y \Gamma_y, 
\end{align}
with PH symmetry (\ref{eq:Dirac-PHS}) and antiunitary symmetry: 
\begin{align}
    \widehat{T2}_y H_{\rm 2d D} (k_x,k_y) \widehat{T2}_y^{-1} = H_{\rm 2d D} (k_x,-k_y).
\end{align}
In this symmetry class, the classifying group is given by $\mathbb{Z}^2$~\cite{shiozaki14}. These integers are characterized by the Chern number $\text{Ch}_1$ and the magnetic winding number $W[2_y]$. The Chern number and magnetic winding number for the Dirac Hamiltonians are calculated by 
\begin{align}
    &c_{ab} = \frac{1}{2i } \text{Tr} [ \Gamma_0 \Gamma_a \Gamma_b] \label{eq:Dirac_Chern} \\
    &w_{g,a} = \frac{1}{2 i} \text{Tr} [ \Gamma_0 \Gamma_a \hat{S}_g],  \label{eq:Dirac_w}
\end{align}
with the chiral operator $\hat{S}_g = e^{i\phi_g} C \hat{g}$ satisfies $\{\hat{S}_g,H_{\rm 3dD} (k_a)\}=0$. Here, $\hat{g}$ is an antiunitary operator. The factor $e^{i\phi_g}$ is chosen such that $\hat{S}_g^2=1$. 
 The Dirac Hamiltonians are constructed as 
\begin{subequations}
    \begin{align}
    &H^{e_1}_{\rm 2d D}  : \ \ (\Gamma_0, \Gamma_x, \Gamma_y) = (\tau_z, \tau_y,\tau_x ), \label{eq:mini-11} \\
    &H^{e_2}_{\rm 2d D}  : \ \ (\Gamma_0, \Gamma_x, \Gamma_y) = (\tau_z, -\tau_y,\tau_x ), \label{eq:mini-12}
\end{align}
\end{subequations}
with $C= \tau_x K$ and $\widehat{T2} = \tau_z K$. Thus, the generators of the classifying group are obtained as
\begin{subequations}
\label{eq:gene_example}
\begin{align}
    &e_1 = (\text{Ch}_1=-1, W[2_y]=1), \\
    &e_2 = (\text{Ch}_1=1, W[2_y]=1), 
\end{align}
\end{subequations}
where we use $c_{xy}$ and $w_{2_y,y}$ with $\hat{S}_{2_y} = \tau_y$. 
An element of the classifying group is then represented by their combinations: $le_1 + m e_2$ ($l,m \in \mathbb{Z}$). Therefore, the Chern number is related to the magnetic winding number. This relationship implies that a single chiral Majorana edge mode must cut across $k_x = 0$ due to PH symmetry, when the open boundary condition in the $y$ direction is imposed. 

In the following, we extend the above argument to the 3D TRSB superconducting states with the magnetic point group symmetry, where we assume that the system is described by the 3D Dirac Hamiltonian,
\begin{align}
    H_{\rm 3d D} = M_0 \Gamma_0 + k_x \Gamma_x + k_y \Gamma_y + k_z \Gamma_z, 
\end{align}
and preserve PH symmetry and magnetic point group symmetry. We focus on the pairing symmetries: $B_{1g} + iB_{1g}$, $A_u + iA_u$, $B_{1u} + i B_{1u}$, $A_u + iB_{1g}$, $B_{2u} + i B_{3u}$, and $B_{1g} + i A_u$. Hereafter, $\sigma_i$, $\tau_i$, and $\eta_i$ denote different Pauli matrices.

\subsection{$B_{1g} + i B_{1g}$ pairing}
First, we consider the symmetry class of $B_{1g} + i B_{1g}$ pairing, whose magnetic point group is
\begin{align}
    mmm = \{e,2_z, 2_y, 2_x, I, m_{z}, m_{y}, m_{x}\}, \label{eq:mmm_elements}
\end{align}
and the pair potential satisfies $\eta_{2_z} = -\eta_{2_y} = -\eta_{2_x} = \eta_{I} = 1$. From Table~\ref{tab:pairings}, $\mathcal{K}_a^{(2)}=\mathbb{Z}^2$. These topological invariants are characterized by $\text{Ch}_1 [m_i]$ in the $m_i \ (i=x,y)$ mirror plane.
The generators of the classifying group are given by
\begin{subequations}
\label{eq:gen_b1g-b1g}
   \begin{align}
    &e_1 = (\text{Ch}_1 [m_x]=-2,\text{Ch}_1 [m_y]=2), \\
    &e_2 = (\text{Ch}_1 [m_x]=2,\text{Ch}_1 [m_y]=2), 
\end{align} 
\end{subequations}
and thus, an element of the classifying group is spanned by the generators, $le_1 + m e_2$ ($l,m \in \mathbb{Z}$).
The matrices of the Dirac Hamiltonian ($\Gamma_0,\Gamma_x,\Gamma_y,\Gamma_z$) are represented as
\begin{subequations}
   \begin{align}
   & H^{e_{1}}_{\rm 3dD} : \ \ (\tau_z,\tau_y \eta_x,\tau_x \eta_x \sigma_z,\tau_y \eta_z \sigma_x ), \\
   & H^{e_{2}}_{\rm 3dD} :  \ \  (\tau_z,\tau_y \eta_x,-\tau_x \eta_x \sigma_z,\tau_y \eta_z \sigma_x ),
\end{align} 
\end{subequations}
with  $C= \tau_x K$, $\hat{2}_z = -i\tau_z \sigma_z$, $\hat{2}_y = -i\tau_z \eta_x \sigma_y$, $\hat{2}_x = -i \eta_x \sigma_x$, and $\hat{I} = -\eta_z \sigma_z$. Using Eqs.~(\ref{eq:mirror-Chern}) and (\ref{eq:Dirac_Chern}), the mirror Chern number is calculated as 
\begin{align}
    c_{m_j,ab} = \frac{c_{ab}^+ - c_{ab}^- }{2}, \label{eq:dirac_mirror_chern}
\end{align}
where $c_{ab}^{\pm}$ is the Chern number defined in the eigenspace of $m_j=\pm i$ in the ($ab$) plane. The generators~(\ref{eq:gen_b1g-b1g}) are calculated from $c_{m_x,yz}$ and $c_{m_y,zx}$ with $\hat{m}_x = -i \eta_y \sigma_y $ and $\hat{m}_y = i \tau_z \eta_y \sigma_x $. 

\subsection{$A_{u} + i A_u$ pairing}
\label{app:au+iau}
 The crystalline symmetry of the $A_u + i A_u$ pairing states is given by Eq.~(\ref{eq:mmm_elements}), and the pair potential satisfies $\eta_{2_z} = \eta_{2_y} = \eta_{2_x} = -\eta_{I} = 1$. The classification reads $\mathcal{K}_a^{(2)} = \mathbb{Z}^3$ and $K_{a}^{(3)} = \mathbb{Z}_2$, which are characterized by $\text{Ch}_1 [m_j]$ and the $\mathbb{Z}_2$ index $\nu_2[I]^{m_j}_+$ of $\kappa[I]^{m_j}_+$ in the $m_j$ mirror plane ($j=x,y,z$), respectively. 
The generators of the classifying group are given by
\begin{subequations}
\label{eq:gen_au-au}
    \begin{align}
    &e_1 = (-1,-1,1,1), \\
    &e_2 = (-1,1,-1,1), \\
    &e_3 = (1,-1,-1,1), 
\end{align}
\end{subequations}
where $e_i = (\text{Ch}_1 [m_x],\text{Ch}_1 [m_y],\text{Ch}_1 [m_z],\kappa[I]^{m_x}_{+})$, and $\kappa[I]^{m_x}_{+} = \kappa[I]^{m_y}_{+} = \kappa[I]^{m_z}_{+} $ for the 3D Dirac Hamiltonian. 
The matrices of the 3D Dirac Hamiltonians are given by
\begin{subequations}
\label{eq:hami_au-au}
   \begin{align}
   & H^{e_{1}}_{\rm 3dD} : \ \  (\tau_z,\tau_x \sigma_z, \tau_y , \tau_x \sigma_x ), \\
   & H^{e_{2}}_{\rm 3dD} : \ \ (\tau_z,-\tau_x \sigma_z, \tau_y , \tau_x \sigma_x ), \\
   & H^{e_{3}}_{\rm 3dD} : \ \ (\tau_z,\tau_x \sigma_z, -\tau_y , \tau_x \sigma_x ), 
\end{align} 
\end{subequations}
with  $C= \tau_x K$, $\hat{2}_z = -i\tau_z \sigma_z$, $\hat{2}_y = -i\sigma_y$, $\hat{2}_x = -i\tau_z \sigma_x$, and $\hat{I} = -\tau_z $. The generators of Eq.~(\ref{eq:hami_au-au}) are calculated from $c_{m_x,yz}$, $c_{m_y,zx}$, and $c_{m_z,xy}$ with $\hat{m}_x = -i  \sigma_x $,  $\hat{m}_y = i \tau_z \sigma_y $, and  $\hat{m}_z = i \sigma_z $, where $\kappa[I]^{m_j}_+$ corresponds to the difference of the number of the occupied bands with $\hat{I}= \pm 1$ at $k_x=k_y=k_z=0$ in the eigenspace of $m_j=+i$. 
Notably, Majorana corner states appear only when the generators are proportional to $(-2,0,0,2)$, $(0,-2,0,2)$, or $(0,0,-2,2)$. Therefore, it always accompanies a Majorana helical hinge state on a certain mirror plane.

\subsection{$B_{1u} + i B_{1u}$ pairing}
For the $B_{1u} + i B_{1u}$ pairing state, the pair potential satisfies $\eta_{2_z} = -\eta_{2_y} = -\eta_{2_x} = -\eta_{I} = 1$. The classification changes to $\mathcal{K}_a^{(2)} = \mathbb{Z}$ and $K_{a}^{(3)} = \mathbb{Z}_2$, which are characterized by $\text{Ch}_1 [m_z]$ and $\nu_2[I]^{m_z}_{\pm}=1 \ (\kappa[I]^{m_z}_{\pm}=2)$ in the $m_z$ mirror plane. 
The generators of the classifying group are given by
\begin{align}
    &e_1 = (\text{Ch}_1 [m_z]=2,\kappa[I]^{m_z}_{+}=2). \label{eq:gene_b1u-biu}
\end{align}
The corresponding 3D Dirac Hamiltonian is given by 
\begin{align}
   & H^{e_{1}}_{\rm 3dD} : \ \ (\tau_z,-\tau_y,\tau_x  \eta_z \sigma_z ,\tau_x \eta_z \sigma_x ),  
\end{align}
with  $C= \tau_x K$, $\hat{2}_z = -i\tau_z \eta_z \sigma_z$, $\hat{2}_y = -i\tau_z \eta_y \sigma_z$, $\hat{2}_x = -i\eta_x$, and $\hat{I} = -\tau_z$. The topological invariants are calculated by $c_{m_z,xy}$ and the difference of the number of the occupied bands with $\hat{I}= \pm 1$ at $k_x=k_y=k_z=0$ in the eigenspace of $m_z=+i$. The generator (\ref{eq:gene_b1u-biu}) implies the coexistence of Majorana hinge and corner states in the same mirror plane. When $2 l e_1 = (4l,0)$ ($l \in \mathbb{Z}$), the Majorana corner states are gapped out, and the Majorana hinge states only appear.

\subsection{$A_u+ i B_{1u}$ pairing}
We consider the symmetry class of $A_{u} + i B_{1u}$ pairing. The crystalline symmetry operators are represented as
\begin{align}
    m'm'm = \{e,2_z, I, m_{z}; T2_x, T2_y,Tm_x,Tm_y\}, \label{eq:m'm'm_elements}
\end{align}
and the pair potential satisfies $\eta_{2_z}=-\eta_{I}=-\eta_{m_z} = 1$. The classification reads $\mathcal{K}_a^{(2)}=\mathbb{Z}$ and $\mathcal{K}_a^{(3)}=\mathbb{Z}^2$, which are described by $\text{Ch}_1 [m_z]$, $W[2_x]$, and $W[2_y]$. Note that there also exist $\text{Ch}_1$, $W[m_x]$, and $W[m_y]$ as the topological invariants for nodal superconducting phases. In order for the 3D Dirac Hamiltonians to be fully gapped, it is necessary to satisfy $\text{Ch}_1 = W[m_x] = W[m_y]=0$, which imposes a constraint on  $\text{Ch}_1$, $W[m_x]$, and $W[m_y]$ as follows. From Eq.~(\ref{eq:mirror-ch}), $\text{Ch}_1 = 0$ yields $\text{Ch}_1[m_z] = \text{Ch}_1^+ = - \text{Ch}_1^-$. In addition, since $[\Gamma(T2_i),D(m_z)]=[\Gamma(Tm_j),D(m_z)]=0$ $(i,j=x,y; i \neq j)$ and $\Gamma(T2_i) = i \Gamma(Tm_j) \tilde{D}(m_z) $, $W[2_i]$ and $W[m_j]$ are related by
\begin{subequations}
    \begin{align}
    &W[2_i] = W_+ + W_-, \\
    &W[m_j] = W_+ - W_-,
\end{align}
\end{subequations}
where $W_{\pm}$ is the magnetic winding number $W[2_i]$ defined in the eigenspace of $\tilde{D}(m_z) = \pm i$. Thus,  $W[m_j] = 0$ yields $W[2_i] = 2 W_+ \in 2 \mathbb{Z}$. With these constraints in mind, the generators of the classifying group are constructed as
\begin{subequations}
    \begin{align}
    &e_1 = (-1,-2,2), \\
    &e_2 = (1,2,2), \\
    &e_3 =  (1,-2,-2),
\end{align}
\end{subequations}
where  $e_i = (\text{Ch}_1 [m_z],W[2_x],W[2_y])$. The matrices of 3D Dirac Hamiltonians for each generator are given by
\begin{subequations}
    \begin{align}
    &H^{e_1}_{\rm 3dD} : \ \ (\tau_z \sigma_z,\tau_y \sigma_z, \tau_x \sigma_z, \sigma_x ), \label{eq:mini-21} \\
    &H^{e_2}_{\rm 3dD} : \ \ (\tau_z \sigma_z,-\tau_y \sigma_z, \tau_x \sigma_z, \sigma_x ), \label{eq:mini-22} \\
    &H^{e_3}_{\rm 3dD} : \ \ (\tau_z \sigma_z,\tau_y \sigma_z, -\tau_x \sigma_z, \sigma_x ), \label{eq:mini-23} 
\end{align}
\end{subequations}
with $C=\tau_x K$, $\widehat{T2}_y=\tau_z K $, $\widehat{T2}_x = K $, and $\hat{m}_z = i \sigma_z$. The topological numbers are calculated by $c_{m_z,xy}$, $w_{2_x,x}$, and $w_{2_y,y}$, where $\hat{S}_{2_x} = \tau_x$ and $\hat{S}_{2_y} = \tau_y$.

\subsection{$B_{2u}+ i B_{3u}$ pairing}
\label{app:B2u+iB3u}
 The crystalline symmetry of the $B_{2u} + i B_{3u}$ pairing states is the same as that of the $A_u+ i B_{1u}$ pairing state [Eq.~(\ref{eq:m'm'm_elements})], whereas the pairing symmetry changes to $-\eta_{2_z}=-\eta_{I}=\eta_{m_z} = 1$. Since $\mathcal{K}_a^{(3)} = \mathbb{Z}_2$, the topological classification predicts the existence of Majorana corner states characterized by the $\nu_3[I] \in \mathbb{Z}_2$. A nontrivial element of $\nu_3[I] \in \mathbb{Z}_2$ is generated by the 3D Dirac Hamiltonian:  
\begin{align}
    &H_{\rm 3dD} : \ \ (\tau_z, \tau_y \eta_x \sigma_x,\tau_x \sigma_x,\tau_x \eta_z \sigma_z ),
\end{align}
with $C= \tau_x K$, $\widehat{T2}_x=K$, $\hat{I} = \tau_z$, and $\hat{m}_z = i \tau_z \sigma_z$. Here, $\nu_3[I]$ is evaluated by the difference of the number of the occupied bands with $\hat{I}= \pm 1$ at $k_x=k_y=k_z=0$.

\subsection{$B_{1g} + i A_u$ pairing}
Finally, we consider the symmetry class of $B_{1g} + i A_u$ pairing states. The crystalline symmetry operators are given by 
\begin{align}
    mmm' = \{e,2_z, m_{x}, m_{y}; TI, T2_x,T2_y,Tm_{z}\},
\end{align}
and the pair potential satisfies $\eta_{2_z}=-\eta_{m_x} = -\eta_{m_y} = 1$. We have $\mathcal{K}_a^{(2)}=\mathbb{Z}^2$ and $\mathcal{K}_a^{(3)}=\mathbb{Z}^2$, which comes from $\text{Ch}_1 [m_x]$ and $W[2_y]$ in the $m_x$ mirror plane and $\text{Ch}_1 [m_y]$ and $W[2_x]$ in the $m_y$ mirror plane, where $\text{Ch}_1 = 0$ due to $TI$ symmetry. The generators of the classifying group are obtained as
\begin{subequations}
   \begin{align}
    &e_1 = (-1,-1,-2,2), \\
    &e_2 = (1,1,-2,2), \\
    &e_3 = (1,-1,-2,-2), \\
    &e_4 = (-1,1,-2,-2),
\end{align} 
\end{subequations}
 where $e_i = (\text{Ch}_1 [m_x],\text{Ch}_1 [m_y],W[2_x],W[2_y])$. The matrices of the 3D Dirac Hamiltonian are given by
\begin{subequations}
 \begin{align}
   & H^{e_{1}}_{\rm 3dD} : \ \ (\tau_z,\tau_y \sigma_x,\tau_x,\tau_y \sigma_z), \label{eq:mini-31} \\
   & H^{e_{2}}_{\rm 3dD} : \ \ (\tau_z,\tau_y \sigma_x,\tau_x,-\tau_y \sigma_z), \label{eq:mini-32} \\
   & H^{e_{3}}_{\rm 3dD} : \ \ (\tau_z,\tau_y \sigma_x,-\tau_x,\tau_y \sigma_z), \label{eq:mini-33} \\
   & H^{e_{4}}_{\rm 3dD} : \ \ (\tau_z,\tau_y \sigma_x,-\tau_x,-\tau_y \sigma_z), \label{eq:mini-34} 
\end{align}   
\end{subequations}
with  $C= \tau_x K$, $\widehat{T2}_y = \tau_z K$, $\widehat{m}_x = i \sigma_z $, and $\widehat{m}_y = i \tau_z \sigma_y $. The topological invariants are calculated from $c_{m_x,yz}$, $c_{m_y,zx}$, $w_{2_x,x}$, and $w_{2_y,y}$, where $\hat{S}_{2_x} = \tau_x \sigma_x$ and $\hat{S}_{2_y}=\tau_y$.

\bibliography{ref} 

\end{document}